\documentclass[conference,compsoc]{IEEEtran}

\usepackage{tikz}
\usepackage{amsmath}

\usepackage{filecontents}

\usepackage{graphicx}
\usepackage{pifont}
\usepackage{xcolor}
\usepackage[ruled,vlined,linesnumbered]{algorithm2e}
\usepackage{amsmath}
\usepackage{amsthm}
\newtheorem{definition}{Definition}
\usepackage{booktabs}
\usepackage{subfigure}
\usepackage{cuted}
\usepackage{hyperref}
\hypersetup{
    colorlinks=true,
    linkcolor=black,
    citecolor=black,
    urlcolor=cyan
}
\usepackage{amsthm}
\usepackage{amssymb}

\begin{document}

\date{}

\title{\Large \bf The Walls Have Ears:\\ Unveiling Cross-Chain Sandwich Attacks in DeFi}

\author{
\IEEEauthorblockN{
Chuanlei Li\textsuperscript{\dag},
Zhicheng Sun\textsuperscript{\dag},
Jing Xin Yuu\textsuperscript{\dag},
Xuechao Wang\textsuperscript{\dag}
}
\IEEEauthorblockA{
\textsuperscript{\dag}The Hong Kong University of Science and Technology (Guangzhou)
}
\IEEEauthorblockA{
Email:\{cli839,zsun277,jxyuu971\}@connect.hkust-gz.edu.cn, xuechaowang@hkust-gz.edu.cn
}
}

\maketitle

\begin{abstract}
Cross-chain interoperability is a core component of modern blockchain infrastructure, enabling seamless asset transfers and composable applications across multiple blockchain ecosystems. However, the transparency of cross-chain messages can inadvertently expose sensitive transaction information, creating opportunities for adversaries to exploit value through manipulation or front-running strategies.

In this work, we investigate cross-chain sandwich attacks targeting liquidity pool-based cross-chain bridge protocols. We uncover a critical vulnerability where attackers can exploit events emitted on the source chain to learn transaction details on the destination chain before they appear in the destination chain mempool. This information advantage allows attackers to strategically place front-running and back-running transactions, ensuring that their front-running transactions always precede those of existing MEV bots monitoring the mempool of the destination chain. Moreover, current sandwich-attack defenses are ineffective against this new cross-chain variant. To quantify this threat, we conduct an empirical study using cross-chain transaction data from the Symbiosis protocol (November 21, 2022 to April 4, 2026) and Debridge (December 21, 2022 to April 4, 2026), along with a tailored heuristic detection model. Our analysis indicates that this attack has indeed had a significant impact on DeFi.
\end{abstract}

\section{Introduction}

Decentralized Finance (DeFi) has emerged as one of the most transformative innovations in the blockchain ecosystem, fundamentally reshaping financial intermediation. The DeFi market has experienced exponential growth, with the total value locked (TVL) across protocols exceeding 145 billion USD as of November 3, 2025~\cite{DeFiLlama}. At the heart of this ecosystem are decentralized exchanges (DEXs), which enable users to trade digital assets directly on-chain without intermediaries. Unlike traditional order-book exchanges, DEXs rely on Automated Market-Making (AMM) mechanisms, where smart contracts autonomously determine exchange rates based on liquidity pool reserves~\cite{zhou2023sok}. Every trade, price update, and liquidity adjustment is immutably recorded on the blockchain, ensuring transparency and verifiability for all participants.


\begin{figure}[t]
    \centering
    \includegraphics[width=1\linewidth]{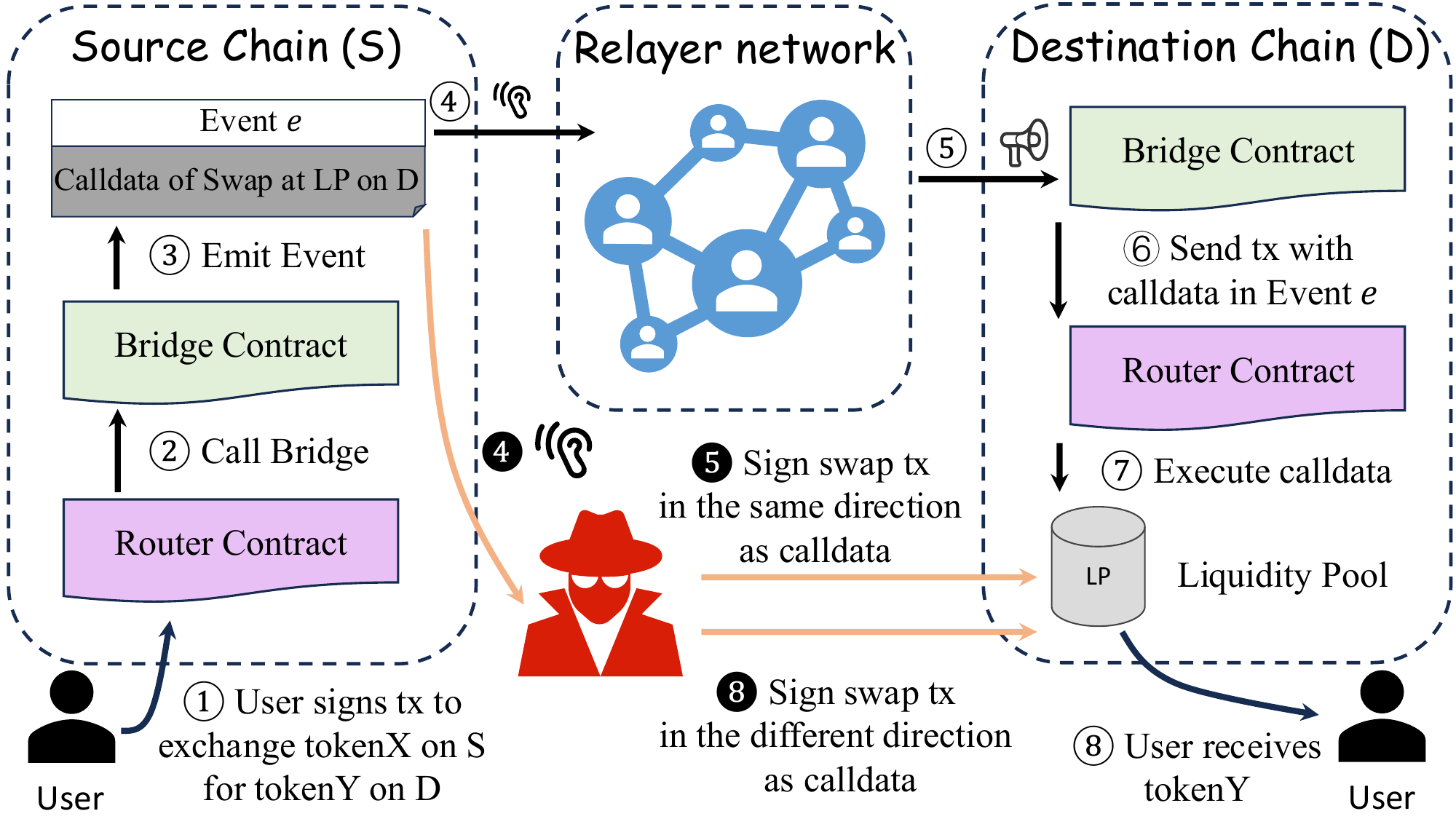}
    \caption{{Overview of the sandwich attack on liquidity pool based cross-chain swaps. \ding{172}-\ding{179} represent normal cross-chain swap steps, while \ding{205},\ding{206},\ding{209} are the attacker's operations. Matching numbers indicate that the user and the attacker perform those steps concurrently.}}
    \label{framework}
\end{figure}

In traditional financial markets, access to transaction information has long been monetized. For example, institutional traders engage in high-frequency trading by minimizing latency to profit from microsecond-level price movements. A similar phenomenon has emerged in DeFi, where participants exploit informational advantages embedded in pending or on-chain transactions to extract profit, a practice broadly known as Maximum Extractable Value (MEV)~\cite{qin2022quantifying}. Various MEV strategies have been proposed and observed. Common MEV strategies include liquidation, arbitrage, and sandwich attacks.
In liquidation, an actor monitors DeFi lending protocols and profitably liquidates positions approaching their collateral thresholds~\cite{qin2021empirical}.
Arbitrage opportunities arise when price discrepancies exist across exchanges or liquidity pools, allowing traders to buy low and sell high, restoring price equilibrium. 
Among these, sandwich attacks~\cite{daian2020flash} are particularly harmful: an attacker observes a victim’s pending transaction in the mempool and strategically places one transaction before it (front-running) and another after it (back-running). By manipulating transaction ordering via gas fees, the attacker inflates the asset price before the victim’s swap and reverses the trade afterward, extracting a profit at the victim’s expense.

Several defense mechanisms have been proposed to mitigate sandwich attacks. 
Proposer-Builder Separation (PBS) mitigates the concentration of MEV by decoupling block construction from block proposal: competitive builders assemble candidate blocks, while validators propose them to the network~\cite{yang2025decentralization,wang2025private}.
Another line of defense involves the encrypted mempool~\cite{choudhuri2024mempool,choudhuri2025practical,bormet2025beat}, where transactions remain encrypted during block construction. Miners or validators build blocks using ciphertexts, and a committee later decrypts the included transactions once the block is finalized, thereby preventing pre-execution visibility and front-running. Furthermore, several works introduce fair transaction ordering at the consensus layer~\cite{kelkar2023themis,li2024sok}, ensuring that users’ transactions are ordered according to predefined fairness principles rather than miners’ discretion, thereby reducing opportunities for manipulation.

While front-running and sandwich attacks have been extensively examined in single-chain DEXs, the rise of cross-chain interoperability introduces a new dimension to these threats. As the DeFi ecosystem expands beyond individual blockchains, users increasingly engage in asset exchanges across multiple chains, relying on cross-chain bridges to facilitate these transfers~\cite{sheng2023trustboost,augusto2024sok}. Such systems enable interoperability by transmitting messages or tokens between heterogeneous blockchain networks, typically through trusted relayers or bridge operators responsible for observing and forwarding cross-chain events~\cite{guo2024zkcross}. However, this process inherently exposes transaction metadata to the public: in EVM-based chains, relayers monitor events emitted by smart contracts, while in UTXO-based chains, they parse newly confirmed transaction outputs. Because these observations occur before the corresponding transactions appear in the destination chain’s mempool, adversaries monitoring the same on-chain events can gain early access to the details. This advantage enables attackers to anticipate and exploit upcoming swaps on the destination chain, paving the way for cross-chain sandwich attacks that are invisible to existing MEV defenses. 


In this paper, we reveal how cross-chain sandwich attacks can be performed on liquidity pool-based bridge protocols. In these protocols, the source chain emits detailed cross-chain swap information, while the destination chain executes the corresponding token swap from a local liquidity pool and sends the output tokens to the user.
As illustrated in Figure~\ref{framework}, an attacker continuously monitors specific events emitted by cross-chain protocols to learn about forthcoming cross-chain swap transactions on the destination chain. Once a profitable opportunity is detected, the attacker submits a front-running transaction to the corresponding liquidity pool on the destination chain and subsequently places a back-running transaction once the victim’s transaction appears on-chain or in the mempool, thereby completing the sandwich attack. Crucially, the attacker acquires the necessary information before the cross-chain transaction appears in any destination-chain context, giving them a temporal advantage over ordinary single-chain sandwich attackers. Moreover, this early access renders most previously proposed mitigations against single-chain sandwich attacks ineffective,as it does not need to wait for victim's transaction to be visible in the destination chain's mempool. We provide an example of a cross-chain sandwich attack that we identified on Ethereum Mainnet (\href{https://etherscan.io/tx/0x9d5f874f8fde164a45cc758611424c6e91e60ed98f265f1a6ed1f11b5482d548}{\textcolor{cyan}{0x9d5f...d548}}, \href{https://etherscan.io/tx/0xef068c4ae7e56a54a72432887dc9faa6ea2f7ce9606d8ddc758c975970692d7a}{\textcolor{cyan}{0xef06...2d7a}}, \href{https://etherscan.io/tx/0xe16b2360053139b1d2991c26380529007ddd20ca6bb8d5ec59076bb4853282c0}{\textcolor{cyan}{0xe16b...82c0}}, \href{https://etherscan.io/tx/0xc07da1dea2dba2baf985e2f774342f3ebdec551d40905bf9400646e04f612b8f}{\textcolor{cyan}{0xc07d...2b8f}}, \href{https://etherscan.io/tx/0xe2a1c948e86bdc4785bf3fee56218a16d823ccbd56662c0d1d490249a94986de}{\textcolor{cyan}{0xe2a1...86de}}), in which the cross-chain attacker achieved a higher profit rate of 21.4\%, compared to just 0.8\% for existing MEV bots. 


We begin by presenting a theoretical framework for analyzing cross-chain sandwich attacks. Since the attack dynamics depend on several stochastic parameters, we extract their empirical distributions from real-world data and verify that the observed values closely match our theoretical assumptions.
Next, we examine historical cross-chain transaction data to identify concrete instances of such attacks. We collect cross-chain transaction records from Symbiosis~\cite{SymbiosisDocumentation} (November 21, 2022 to April 4, 2026) and deBridge~\cite{deBridge} (December 21, 2022 to April 4, 2026), then apply our heuristic detection algorithm to systematically locate cross-chain sandwich behaviors. Our analysis reveals that these attacks generated a profit of approximately 0.56 million USD, representing 1.75\% of the total bridged volume during the observation period. By contrast, conventional single-chain sandwich attacks in the same dataset yielded only 6475.91 USD in profit. Furthermore, we found that cross-chain sandwich attacks have the greatest impact on Ethereum. Among the transactions in our dataset, those targeting Ethereum as the destination chain generated the highest profits, with the StableCoin\(\leftrightarrow\)WETH pools being the most frequently attacked.
Finally, we discuss potential countermeasures and outline design principles aimed at mitigating this emerging class of cross-chain threats.

\textbf{Contributions.} We summarize our contributions as follows.

\begin{itemize}
    \item We identify a previously unrecognized cross-chain sandwich attack that exploits information advantage in cross-chain protocols. By accessing transaction details before they appear on the destination chain, the attacker can consistently outperform existing MEV bots in transaction-ordering competitions, effectively bypassing all known single-chain defenses.
    \item We conduct a large-scale empirical study using historical cross-chain transaction data, revealing that attackers extracted approximately 0.56 million USD in profit through such attacks, equivalent to 1.75\% of the total bridged volume.
    \item To foster further research, we construct and release the first large-scale dataset\footnote{\href{https://www.kaggle.com/datasets/anonymous8397458/cross-chain-sandwich-datasets}{\textcolor{black}{https://www.kaggle.com/datasets/anonymous8397458/cross-chain-sandwich-datasets}}} of cross-chain sandwich attacks, providing an empirical foundation for future work on cross-chain MEV and security analysis.
\end{itemize}

\section{Background}
In the following, we provide relevant background on DEX, sandwich attack, and cross-chain interoperability.

\subsection{Decentralized Exchange}
Unlike centralized exchange (CEX), which requires users to deposit funds and trust the centralized entity to manage order books and execute trades, Decentralized Exchange (DEX) allows users to trade cryptocurrencies directly from their personal wallets without the need for an intermediary.

DEX employs Automated Market Maker (AMM) instead of traditional order books. In this model, liquidity providers (LPs) supply assets to liquidity pools that contain at least two tokens. The price of a token is then determined algorithmically based on the asset ratio in the pool, according to rules defined in an immutable and transparent smart contract. A classic AMM model is the Constant Product Market Maker (CPMM), used in Uniswap V2~\cite{Uniswap}. It is also the model leveraged in Uniswap V3 when the price does not cross a tick, which is the case for almost all trades. For the reserves \(x\) and \(y\) in the liquidity pool, their quantities always satisfy the equation 
\begin{equation}
    x*y=k.
\end{equation}
We consider that a user aims to swap \(\Delta x\) of token X to token Y in a liquidity pool with transaction fee \(f\in(0,1)\), the reserves follow this equation:
\begin{equation}
    (x+(1-f)\Delta x)(y-\Delta y)=x*y,
\end{equation}
the user receives 
\begin{equation}
\scalebox{0.99}{$
    \Delta y=y-\frac{x*y}{x+(1-f)\Delta x}=\frac{y(1-f)\Delta x}{x+(1-f)\Delta x}=\frac{k(1-f)\Delta x}{x^2+x(1-f)\Delta x},
    $
    }
\end{equation}
and 
\begin{equation}
    \frac{\partial\Delta y}{\partial x}=-\frac{k(2x+(1-f)\Delta x)(1-f)\Delta x}{(x^2+x(1-f)\Delta x)^2}<0.
\end{equation}
So an increase in reserve \(x\) leads to a decrease in \(\Delta y\). This implies that large transactions can cause price deviations, a phenomenon known as slippage. A slippage tolerance \(s\in(0,1)\) represents the difference between the expected out \(\Delta y\) and actual execution output \(\Delta y'\) of a transaction, where \(\Delta y'\geq(1-s)\Delta y\). If the actual execution amount falls below the minimum acceptable value, the transaction will be reverted.

\subsection{Sandwich Attack}
\label{sandwiching}
The slippage tolerance creates an opportunity for predatory traders to perform their strategies to extract value. The attack exploits the public nature of the mempool, which contains unconfirmed transactions broadcast to the network. Attackers monitor the mempool to identify profitable pending transactions targeting specific liquidity pools. They then bid higher gas fees to get miners to include a same-direction transaction (front-running) before the victim’s transaction. This front-running raises the token price, so the victim receives fewer tokens than expected. Next, the attacker submits an opposite-direction transaction (back-running) with a slightly lower gas fee, ensuring it is placed immediately after the victim's transaction. Finally, the attacker trades the overpriced token back and captures the profit. 

Consider a pool with reserves \((x_0,y_0)\) for tokens X and Y, and a transaction fee rate \(f\in(0,1)\). At time \(t_0\), the attacker identifies a victim transaction \(T_v\).
\(T_{v}\) aims to exchange \(\Delta x_v\) X for Y with slippage tolerance \(s\in(0,1)\). If \(T_v\) is not attacked, it will return 
\begin{equation}
    \Delta y=\frac{y_0(1-f)\Delta x}{x_0+(1-f)\Delta x}.
\end{equation}

If an attacker is trying to attack \(T_v\) at \(t_0\). Attacker executes a front-running transaction \(T_{A1}\) to input \(\Delta x_{A_1}\) X in the pool, and it will return 
\begin{equation}
    \Delta y_{A_1}=\frac{y_0(1-f)\Delta x_{A_1}}{x_0+(1-f)\Delta x_{A_1}}.
\end{equation}

After \(T_{A1}\), the pool reserves update to \(x_{A_1}=x_0+\Delta x_{A_1}\) and \(y_{A_1}=y_0-\Delta y_{A_1}=\frac{x_0y_0}{x_0+(1-f)\Delta x_{A_1}}\). Then \(T_v\) is executed, the output of Y should be 
\begin{equation}
    \Delta y_v=\frac{y_{A_1}(1-f)\Delta x_v}{x_{A_1}+(1-f)\Delta x_v}.
\end{equation}

If \(\Delta y_v\geq(1-s)\Delta y\), the execution will be successful, otherwise it will be reverted so that the sandwich attack fails. According to the theorem proved by Heimbach et al.~\cite{heimbach2022eliminating}, the attacker's profit-maximizing optimal input amount \(\Delta x_{A1}\) is the value that satisfies \(\Delta y_v=(1-s)\Delta y\), i.e.,
\begin{equation}
\label{cc}
    \frac{\frac{x_0y_0}{x_0+(1-f)\Delta x_{A_1}}(1-f)\Delta x_v}{x_0+\Delta x_{A1}+(1-f)\Delta x_v}=(1-s)\frac{y_0(1-f)\Delta x}{x_0+(1-f)\Delta x}.
\end{equation}

To retrieve the profit, the attacker places the back-running transaction \(T_{A2}\) closely following \(T_{v}\), it exchanges \(\Delta y_{A_1}\) Y with pool reserves \(x_v=x_{A_1}+\Delta x_v\) and \(y_v=y_{A_1}-\Delta y_v=\frac{x_{A_1}y_{A_1}}{x_{A_1}+(1-f)\Delta x_v}\). In transaction \(T_{A2}\), attacker will receive
\begin{equation}
    \Delta x_{A_2}=\frac{x_v(1-f)\Delta y_{A_1}}{y_v+(1-f)\Delta y_{A_1}}.
\end{equation}

Let \(G_c\) be the total gas cost for \(T_{A1}\) and \(T_{A2}\), the profit \(P\) of this sandwich attack is 
\begin{equation}
\label{profitequ}
    P=\Delta x_{A_2}-\Delta x_{A_1}-G_c.
\end{equation}

\subsection{Cross-Chain Interoperability}
Cross-chain interoperability refers to the ability of distinct blockchain networks to reliably transfer or exchange data and assets, enabling seamless interaction among otherwise isolated systems. It has become a foundational component of modern multi-chain ecosystems, allowing users and applications to move liquidity, coordinate state, and compose protocols across heterogeneous chains.

One early interoperability approach is the Lock-and-Mint model~\cite{tairi2023ledgerlocks}: assets are locked on the source chain, and an equivalent amount of wrapped tokens is minted on the destination chain, ensuring the original assets cannot be double-spent.
The reverse process, Burn-and-Unlock~\cite{chainlink}, burns the wrapped tokens on the destination chain and unlocks the corresponding assets on the source chain, keeping the total supply consistent across chains.
Another widely studied interoperability mechanism is the cross-chain atomic swap, which guarantees that a cross-chain token swap between two parties either executes fully on both chains or does not occur at all. Atomicity is enforced through Hashed Time-Locked Contracts (HTLCs)~\cite{herlihy2018atomic}, which coordinate conditional execution across chains with cryptographic timeouts.

\begin{figure}[htbp]
    \centering
    \includegraphics[width=\linewidth]{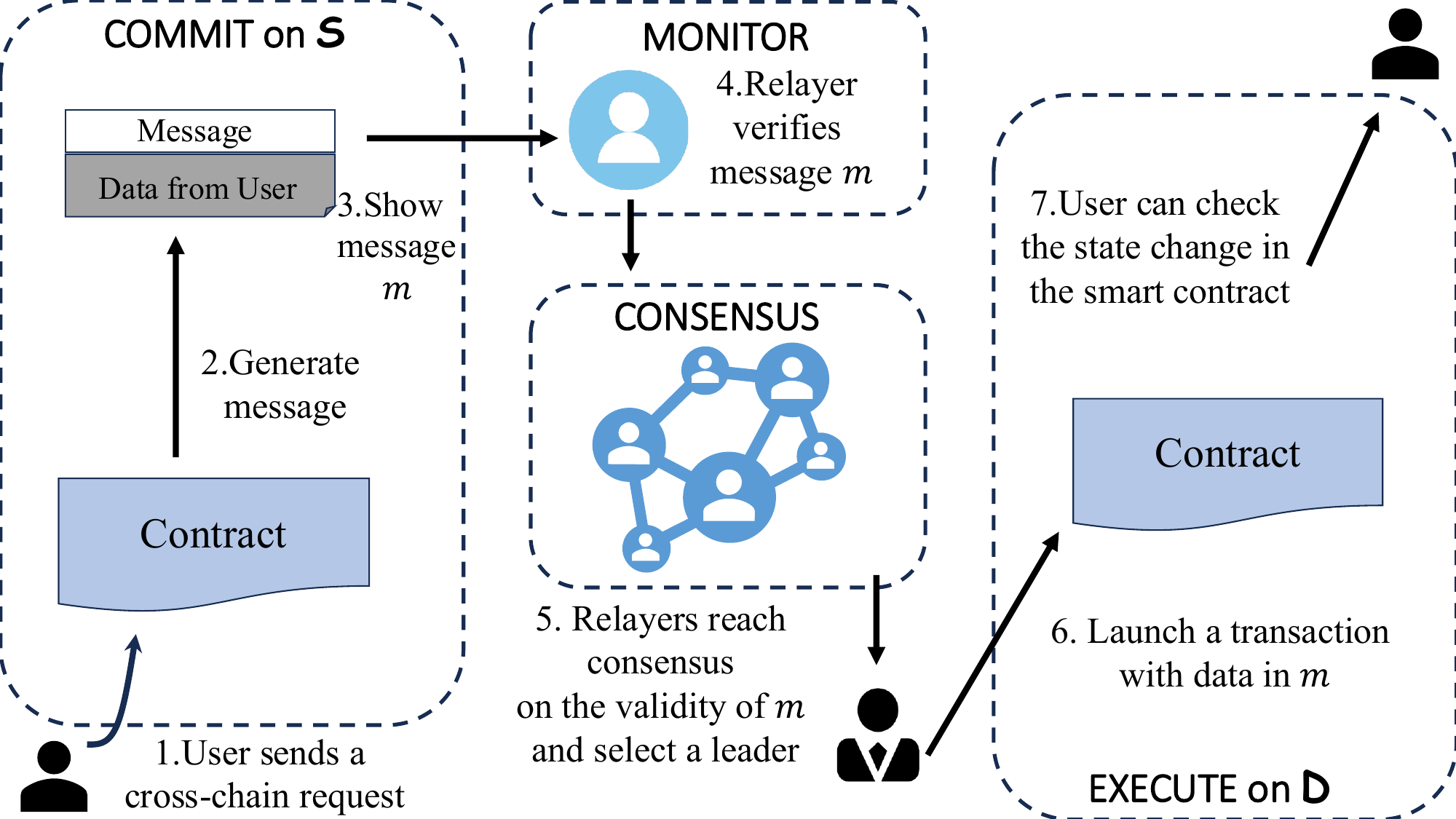}
    \caption{{The workflow of the cross-chain messaging protocol.}}
    \label{ccmp}
\end{figure}

While these mechanisms support basic asset transfers, they fall short when applications require sending arbitrary data across chains. As cross-chain use cases expand, general-purpose cross-chain messaging protocols have become essential, offering greater flexibility and a significantly improved user experience. As illustrated in Figure~\ref{ccmp}, such protocols allow applications to transmit structured messages rather than only wrapped tokens, enabling richer interactions across heterogeneous chains.
Several notable designs exist. Chainlink’s Cross-Chain Interoperability Protocol (CCIP)~\cite{CCIP} provides a generalized messaging framework built atop decentralized oracle networks and a risk-management layer. Cosmos IBC~\cite{cosmosibc} instead uses on-chain light clients so that two blockchains can mutually verify each other’s state through Merkle commitments, avoiding trusted intermediaries altogether.
We define a cross-chain messaging protocol as follows:


\begin{definition}[Cross-Chain Messaging Protocol]
    Source chain \(\mathcal{S}\), destination chain \(\mathcal{D}\), relayers \(\mathcal{R}\), users \(\mathcal{U}\) and contract \(\mathcal{C}\) engage in this protocol to transfer messages across different chains. The contracts \(\mathcal{C}\) are deployed on multiple chains. The cross-chain messaging protocol (CCMP) has the following sub-protocols:

\[
{\fontsize{10}{10}\selectfont CCMP=(\textsc{Commit},\textsc{Verify},\textsc{Consensus},\textsc{Execute})}
\]
    
    \begin{itemize}
        \item \(\textsc{Commit}(\mathcal{S},\mathcal{C},\mathcal{U},\mathcal{R})\rightarrow m\): \(\mathcal{U}\) send a request to the contract \(\mathcal{C}\) on \(\mathcal{S}\), including the data they want to transfer. \(\mathcal{C}\) generates the message \(m\) after encoding the data and emits \(m\) to relayers \(\mathcal{R}\).
        \item \(\textsc{Verify}(m,\mathcal{R})\rightarrow (0,1):\) This protocol is executed by every relayer in \(\mathcal{R}\). After receiving \(m\), they verify its validity. If \(m\) is valid, it outputs 1.
        \item \(\textsc{Consensus}(m,\mathcal{R},\mathcal{D})\rightarrow l:\) All relayers participate in the consensus to determine the validity of \(m\). If the majority considers \(m\) valid, a leader \(l\) is selected to perform \(\textsc{Execute}\) protocol.
        \item \(\textsc{Execute}(l,m,\mathcal{D},\mathcal{C},\mathcal{U})\rightarrow out:\) The leader \(l\) parses the message \(m\) and gets the data from \(\mathcal{U}\). Then \(l\) calls \(\mathcal{C}\) to sign the transaction corresponding to the data. At last, user can verify the cross-chain message by checking the transaction execution result.
    \end{itemize}
\end{definition}

To support decentralization, many cross-chain messaging protocols rely on events emitted by smart contracts to serve as the message \(m\), which permissionless relayers then observe and forward. Because these events are publicly visible and cross-chain execution introduces inherent delays, adversaries can access sensitive transaction information well before it is executed on the destination chain. This early visibility creates a predictable window for MEV extraction. Among all known MEV strategies, sandwich attacks are particularly harmful because they directly degrade user execution quality through price manipulation.
As the pioneering work to study MEV in cross-chain settings, we focus specifically on cross-chain sandwich attacks, which arise in messaging protocols that integrate liquidity-pool swaps on the destination chain. In such protocols, a cross-chain sandwich vulnerability exists when the following properties hold:
\begin{itemize}
    \item The source-chain message becomes publicly observable before the destination-chain swap is executed.
    \item The message reveals, or allows an observer to infer, enough information about the future swap, such as the destination-chain, AMM pool, swap direction, input amount, and slippage bound.
    \item The destination-chain swap is executed against a public liquidity pool whose state can be changed by other swap transactions.
    \item The relay latency between the start transaction on the source chain and the end transaction on the destination chain is long enough for an adversary to submit at least one transaction to the relevant pool.
\end{itemize}

When these conditions hold, attackers can observe the emitted event on the source chain and predict the exact swap that will soon occur on the destination chain, well before the transaction reaches the local mempool. This advance knowledge allows attackers to reliably construct profitable sandwich transactions, a behavior that we have observed in the wild.

Prominent cross-chain messaging protocols with this vulnerability include ThorSwap~\cite{ThorSwap}, deBridge~\cite{deBridge}, and Symbiosis~\cite{SymbiosisDocumentation}, which have processed total cross-chain volumes of 114.15 billion, 23.6 billion, and 9.7 billion USD respectively. Such scale underscores the importance of analyzing and mitigating these emerging cross-chain threats.

\section{Cross-Chain Sandwich Attack}
\label{attacker model}


In this section, we introduce cross-chain sandwich attacks. Figure~\ref{txorder} illustrates all possible transaction orderings, including the cases of no attack, single-chain sandwich attacks only, and cross-chain sandwich attacks, both in isolation and when coexisting with a single-chain sandwich attack. Table~\ref{table:symbols} summarizes all notation used throughout the analysis.

\begin{table}[t]
\caption{Summary of Notations}
\label{table:symbols}
\centering
\scalebox{0.87}{
\begin{tabular}{cl}
\toprule
\textbf{Symbol} & \textbf{Description} \\ 
\midrule

\(\mathcal{A}\)      & Attacker that performs cross-chain sandwich attack \\
\(\mathcal{B}\)      & Attacker that performs single-chain sandwich attack\\
\( f \)         & Swap fee rate of the pool which is in \((0,1)\)\\
\(T_s\)         & Start transaction on the source chain\\
\( T_v \)         & Victim's transaction on the destination chain\\
\(\pi\)          & Representing symbols \(A1,A2,v,B1,B2\) \\& or numbers \(1...n\)\\
\( s_\pi \)         & Slippage tolerance of \(T_\pi\) which is in \((0,1)\)\\
\(N_\pi\)            &Block number of transaction \(T_\pi\)\\
\(T_{A1} \)        & Front-running transaction from \(\mathcal{A}\) \\
\( T_{A2} \)         & Back-running transaction from \(\mathcal{A}\)\\
\( T_{B1}\)          & Front-running transaction from \(\mathcal{B}\) \\
\( T_{B2} \)       & Back-running transaction from \(\mathcal{B}\) \\
\( (x_0,y_0) \)       & Reserves of token X and Y before \(T_{A1}\) \\
\((x_\pi,y_\pi)\)        & Reserves of token X and Y after \(T_\pi\)\\
\((\Delta x_\pi,\Delta y_\pi)\)        & Amounts of token X and Y exchanged in \(T_\pi\)\\
\(p\)        & Probability that \(\mathcal{A}\) does not incur a loss after the \\& execution of \(T_v\) with noisy transactions.\\
\(q\)        & Probability that there is no noisy transactions \\&between \(T_{A1}\) and \(T_v\)\\
\(\theta\)        & Percentage that \(\mathcal{A}\) expects to extract from \(T_v\)\\
\bottomrule
\end{tabular}
}

\end{table}

\subsection{Adversarial Model}
We model the adversary \(\mathcal{A}\) as a rational, profit-driven MEV searcher with no privileged control over the cross-chain protocol. \(\mathcal{A}\) can use RPC services on both \(\mathcal{S}\) and \(\mathcal{D}\), continuously monitor public source-chain events, decode calldata with open-source logic, and locally simulate contract execution when calldata formats are not directly documented. After learning a profitable future swap, \(\mathcal{A}\) can submit ordinary transactions to AMM pools on \(\mathcal{D}\), choose gas prices, and monitor the execution of \(T_v\) to submit a back-running transaction. \(\mathcal{A}\) has sufficient cryptocurrency in its wallet to perform the front-running swap and behaves strategically: it attacks only when the estimated profit exceeds gas cost and AMM fees.

\subsection{Attack Overview}
\(\mathcal{A}\) monitors the source-chain message \(m\) emitted during \(\textsc{Commit}\). If \(m\) reveals enough information to reconstruct the future destination-chain swap \(T_v\), \(\mathcal{A}\) evaluates whether putting swap transaction before \(T_v\) is profitable. When the opportunity is profitable, \(\mathcal{A}\) submits a same-direction transaction \(T_{A1}\) on \(\mathcal{D}\) before \(T_v\) is executed. This front-running transaction moves the AMM price against the victim while keeping the expected output of \(T_v\) above its minimum acceptable amount, and later submits an opposite-direction transaction \(T_{A2}\) after \(T_v\) to unwind the position and realize the price difference.

\begin{figure}[htbp]
    \centering
    \includegraphics[width=\linewidth]{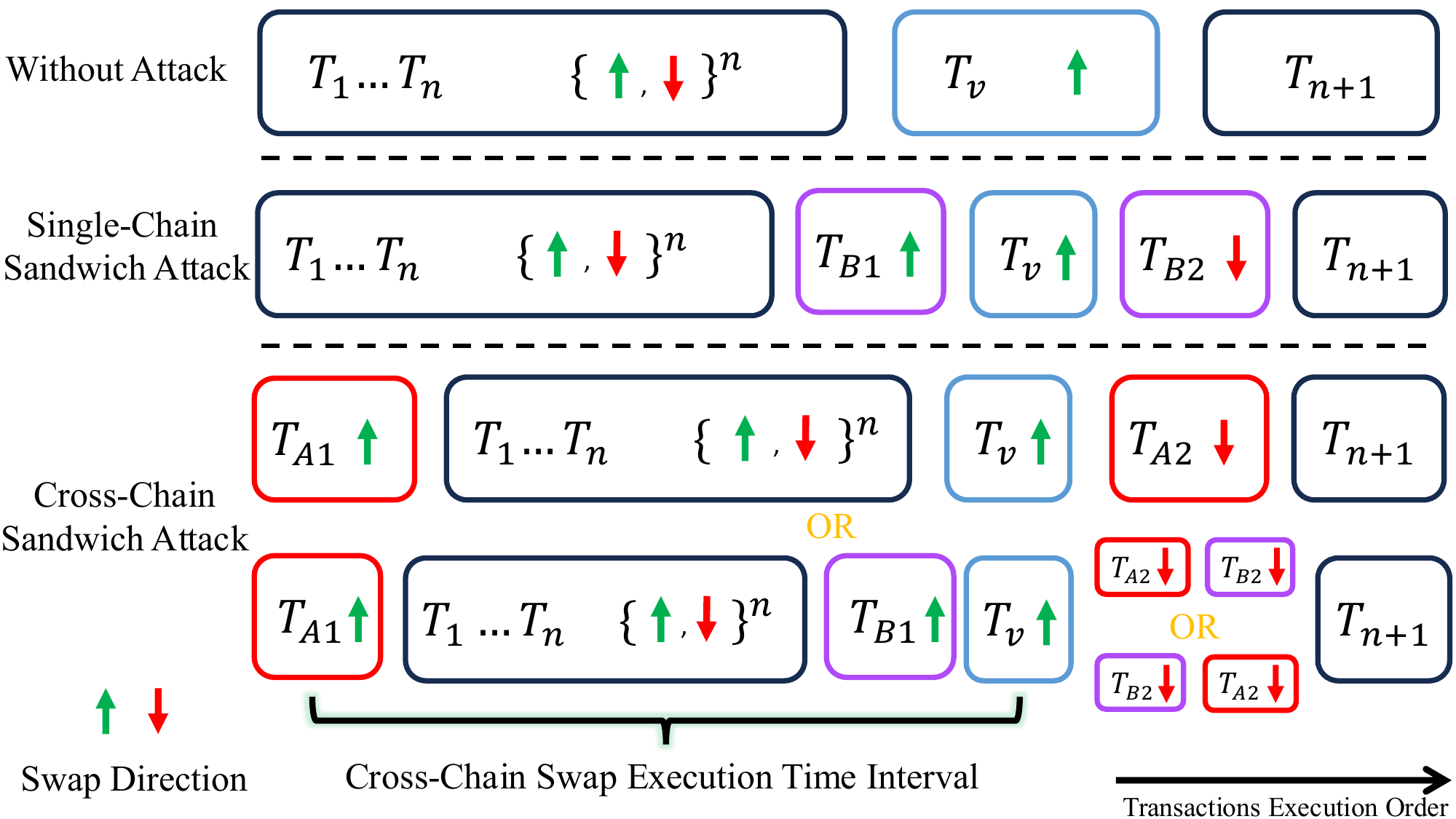}
    \caption{{Visualization of possible transaction orderings. \(T_v\) is the victim transaction. \(T_{A1}\) and \(T_{A2}\) are cross-chain sandwich attack transactions. \(T_{B1}\) and \(T_{B2}\) are single-chain sandwich attack transactions. \(T_1-T_{n}\) and \(T_{n+1}\) are other noisy transactions with unknown swap directions.}}
    \label{txorder}
\end{figure}

\subsection{Noisy Transactions}
It is worth noting that between the event being emitted on the source chain and the swap being executed on the destination chain, the message must be relayed by at least one intermediary (\ding{176}\ding{177}\ding{178} in Figure~\ref{framework}). During this time interval, numerous swaps in unknown directions may occur in the corresponding liquidity pool, which is reflected as \(T_1...T_{n}\) in the figure.
We call these swaps noisy transactions and denote \(q\) as the probability that no noisy transactions exist between \(T_{A1}\) and \(T_v\). Noisy transactions may change the pool’s state to one that does not meet \(\mathcal{A}\)’s expectations, thereby directly affecting the potential profit of its cross-chain sandwich attack on \(T_v\). As an extreme case, if a large-volume transaction in noisy transactions occurs in the direction opposite to \(T_{A1}\), then even if \(T_v\) executes successfully, \(\mathcal{A}\) can only swap back fewer tokens than \(\Delta x_{A_1}\), leading to a loss in its sandwich attack.

As it is impossible to acquire or predict any information about noisy transactions at the time when the event is observed, we regard noisy transactions as a black-box transformation \(f:(x_{A_1},y_{A_1})\rightarrow(x_{n},y_{n})\), where the input and output represent the pool states after \(T_{A1}\) and \(T_{n}\). If \(T_v\) is exchanging X for Y, we denote \(p\) as the probability that \((x_{n}+\Delta x_v,y_{n}-\Delta y_v)\) can ensure that \(\mathcal{A}\) does not incur a loss \((\Delta x_{A2}\geq\Delta x_{A1})\) when \(T_1...T_{n}\) are not empty, where \(\Delta x_{A2}\) is the output amount in the back-running trade \(T_{A2}\) and \(\Delta x_{A1}\) is the input amount in the front-running trade \(T_{A1}\).  

As \(p\) and \(q\) are derived from this empirical, black-box model, we estimate their values through large-scale historical data analysis rather than theoretical derivation. The resulting estimates are presented in the experiment Section~\ref{result}.

In practical settings, leaving a small buffer for \(T_v\) is a reasonable strategy for \(\mathcal{A}\) to prevent noisy transactions from pushing \(T_v\) to its slippage limit and causing a transaction revert. Therefore, when \(T_v\) is about to be executed, there may still exist exploitable opportunities, allowing single-chain attackers \(\mathcal{B}\) to perform sandwich attacks on \(T_v\). We denote by \(\theta \in (0,1)\) the fraction extracted by \(\mathcal{A}\). This leaves a buffer of \((1-\theta)s_v\Delta y\) in token Y, and the victim will expect to receive \((1-\theta)s_v\Delta y\) more than the baseline amount \((1 - s_v)\Delta y\).
 Now the condition in Equation~\ref{cc} is tightened to 
\begin{equation}
\Delta y_v=(1-s_v)\Delta y+(1-\theta)s_v\Delta y=(1-s_v*\theta)\Delta y.
\end{equation}


\subsection{Profit of the Attack}
A direct implication of such attack under our assumptions is that the maximum profit in token X of \(\mathcal{A}\) is \(s_v\Delta x_v\). This value corresponds to the maximum tolerable loss of \(T_v\), and it is independent of other transactions occurring between \(T_{A1}\) and \(T_v\). This is because the minimum output amount of \(T_v\) \(minReturnAmount\) is predefined in the emitted event, i.e., 
determined at the moment \(T_{A1}\) is placed. As long as \(T_v\) can be successfully executed, the corresponding token price in the pool at that time will not exceed the acceptable range of \(T_v\), thus the profit in token X of \(T_{A1}\) can not exceed \(s_v\Delta x_v\). 

We now analyze the practical scenarios shown in Figure~\ref{txorder}, assuming \(T_v\) is trying to exchange \(\Delta x_v\) X for Y. There are four possible scenarios for \(\mathcal{A}\):

\noindent\textbf{No single-chain sandwich attack:} If there is no single-chain sandwich attack, the execution of a cross-chain sandwich attack is the same as that of a single-chain sandwich attack, where the transaction ordering is \((T_{A1},T_1...T_{n},T_v,T_{A2})\). The maximum potential extracted value is \(s_v\Delta x_v\), \(\mathcal{A}\) successfully attacks if \(s_v\Delta x_v>G_c+(1-(1-f)^2)\Delta x_{A_1}\). To show the expected profit, we assume that when the profit is positive, the average profit rate \((\frac{\Delta x_{A2}-\Delta x_{A1}}{\Delta x_{A1}})\) is \(r^+\), and when the profit is negative, the average profit rate is \(r^-\). Considering the parameters \(p\) and \(q\) in noisy transactions, the expectation of profit is 
\begin{equation}
\label{e}
    \mathbb{E}(P)=\Delta x_{A1}((q+(1-q)p)r^+ + (1-q)(1-p)r^-).
\end{equation} 

\noindent\textbf{\(\boldsymbol{T_{A2}}\) executed before \(\boldsymbol{T_{B2}}\):} Although \(\mathcal{A}\) is always able to place \(T_{A1}\) ahead of \(T_{B1}\), it can not guarantee winning the race when inserting back-running transactions. If \(\mathcal{A}\) wins this race, the transaction ordering is \((T_{A1},T_1...T_{n},T_{B1},T_v,T_{A2})\). We can treat \(T_{B1}\) as one of \(T_1...T_{n}\), and the maximum potential extracted value and expectation of profit are the same as that in the situation with no single-chain sandwich attack.

\noindent\textbf{\(\boldsymbol{T_{A2}}\) executed after \(\boldsymbol{T_{B2}}\):} If \(\mathcal{A}\) loses the race, the transaction ordering is \((T_{A1},T_1...T_{n},T_{B1},T_v,T_{B2},T_{A2})\). Because \(\mathcal{B}\) captures the buffer’s surplus, \(\mathcal{A}\)’s maximum attainable profit at this point is \(\theta s_v\Delta x_v\), and the expectation is the same as Equation~\ref{e}.

\noindent\textbf{\(\boldsymbol{T_{A1}}\) is attacked by \(\boldsymbol{\mathcal{B}}\):} If \(\mathcal{A}\) is attacked by \(\mathcal{B}\), the transaction ordering is \((T_{B1}, T_{A1},T_{B2})\). In each of the three scenarios described above, \(\mathcal{A}\)'s profit decreases by at most \(s_{\mathcal{A}}\Delta x_{A1}\) respectively. However, \(\mathcal{A}\) can mitigate the risk of the sandwich attack from \(\mathcal{B}\) through various methods, such as decreasing \(s_{\mathcal{A}}\) or leveraging a private mempool. Thus, this scenario is regarded as having no impact on \(\mathcal{A}\)'s profit.


\subsection{Against Current Countermeasures}

Existing defenses against sandwich attacks are designed for single-chain settings and therefore cannot address attacks that exploit information leakage across chains. In all cases, their protection applies only after a transaction reaches the destination chain, whereas cross-chain sandwich attacks arise before the transaction ever appears in the destination-chain mempool or block-construction pipeline.

Proposer Builder Separation (PBS)~\cite{yang2025decentralization} restricts the proposer’s ability to reorder transactions after seeing their plaintext contents. It ensures that once builders submit sealed blocks, the proposer cannot manipulate the intra-block ordering for MEV extraction. Fair ordering consensus~\cite{kelkar2020order,kelkar2022order,kelkar2023themis} enforce partial transaction-ordering guarantees at the consensus layer by ensuring that transactions are ordered roughly in the sequence they are observed by the network. Both mechanisms only regulate the ordering of transactions that are already visible to the destination-chain nodes. Since a cross-chain victim transaction is predicted from source-chain events before it ever enters the destination chain’s network, these guarantees provide no protection.

Similarly, private or encrypted mempools~\cite{ding2025asymmetric,choudhuri2024mempool,choudhuri2025practical} hide pending transactions after they reach the destination chain. However, in liquidity-pool-based cross-chain bridges, the essential details of the victim transaction are included in source-chain events that must be publicly emitted for relayers to function. Since the leak occurs before any mempool logic applies, hiding mempool contents does not stop the attack.

\begin{algorithm}[!htbp]
    \caption{Detect Sandwich Attacks in Cross-Chain Swaps}
    \label{detect_algo}

    \KwIn{
        Start transaction on the source chain \(T_s\);
        victim transaction on the destination chain \(T_v\);
        detection range of block numbers after \(T_v\), \(num\)
    }

    \KwOut{
        Detected sandwich attack pairs \(pairs\), including all
        \(\{T_{A1},T_v,T_{A2}\}\)
    }

    \BlankLine

    Query the timestamp \(t_s\) of \(T_s\), the block number \(N_v\) of \(T_v\), and map \(t_s\) to the corresponding block \(N_s\) on the destination chain\;

    Get \(pools\) used in \(T_v\);

    \ForEach{\(pool \in pools\)}{
        Query the Swap log \(L_v\) in \(pool\) that belongs to \(T_v\),
        and read its direction \(d\);

        Query all Swap logs \(FrontLogs\) with direction \(d\) in
        \(pool\) between \(N_s\) and \(N_v\) on the destination chain, query all Swap logs \(BackLogs\) with direction \(-d\) in \(pool\) between \(N_v\) and \(N_v+num\) on the destination chain;

        \ForEach{\(L_f \in FrontLogs\)}{
            \(similarAmountTx \leftarrow \emptyset\);

            Read \(tokenInAmount\) and \(tokenOutAmount\) from \(L_f\), query the top-level transaction \(T_f\) containing \(L_f\);

            \ForEach{\(L_b \in BackLogs\)}{
                Read \(tokenInAmount'\) and \(tokenOutAmount'\) from \(L_b\);

                \If{
                    \(\dfrac{tokenInAmount'}{tokenOutAmount}
                    \in [0.9,1.1]\)
                }{
                    \(similarAmountTx.append(L_b)\);
                }
            }

            \(sameFromTx \leftarrow \emptyset\);
            \(sameContractTx \leftarrow \emptyset\);

            \ForEach{\(L_b \in similarAmountTx\)}{
                Query the top-level transaction \(T_b\) containing \(L_b\);

                \eIf{\(T_f.from = T_b.from\)}{
                    \(sameFromTx.append(L_b)\);
                }{
                    \If{
                        \(T_f.from \neq T_b.from\)
                        \textbf{and}
                        \(T_f.to = T_b.to\)
                        \textbf{and}
                        \(\textsc{IsContractAddr}(T_f.to)\)
                    }{
                        \(sameContractTx.append(L_b)\);
                    }
                }
            }

            \eIf{\(sameFromTx \neq \emptyset\)}{
                \(targetBackLog \leftarrow sameFromTx[0]\);
            }{
                \eIf{\(sameContractTx \neq \emptyset\)}{
                    \(targetBackLog \leftarrow sameContractTx[0]\);
                }{
                    \textbf{continue};
                }
            }

            \(pairs.append(\{L_f,L_v,targetBackLog\})\);

            \(BackLogs.remove(targetBackLog)\);
        }
    }
\end{algorithm}

\section{Empirical Analysis}
In the following, we investigate the scale of such cross-chain sandwich attack by analyzing historical transaction data from two cross-chain protocols in order to assess its feasibility and potential impact on the DeFi market.

\subsection{Detection Algorithm}

Our cross-chain sandwich attack detection algorithm is inspired by the method to identify potential single-chain sandwich attacks provided by Qin et al.~\cite{qin2022quantifying}. We employ the following heuristics to detect historical cross-chain sandwich attacks. The pseudocode is provided in Algorithm~\ref{detect_algo}.

\begin{itemize}
    \item \textbf{Direction:} Both \(T_{A1}\) and \(T_v\) exchange token X for token Y, while \(T_{A2}\) exchange token Y for token X.
    \item \textbf{Pair:} Each \(T_{A1}\) can only match one \(T_{A2}\).
    \item \textbf{Amounts:} Consider a perfect sandwich attack, the amount of bought token in \(T_{A1}\) should be equal to the amount of sold token in \(T_{A2}\). In our practice, we relax this constraint by requiring that the amount of sold token in \(T_{A2}\) is between 90\% and 110\% of the amount of bought token in \(T_{A1}\). 
    \item \textbf{Address:} Either the same user address sends transactions \(T_{A1}\) and \(T_{A2}\), or two different user addresses send \(T_{A1}\) and \(T_{A2}\) to the same contract address. In Algorithm~\ref{detect_algo}, we first filter back transactions that satisfy the other heuristics. Among these transactions, the transaction with the same tx.from address as \(T_{A1}\) has higher priority to pair with \(T_{A1}\). These heuristics above we adopt are exactly the same as those proposed by Qin et al.~\cite{qin2022quantifying}.
    \item \textbf{Position:} \(T_{A1}\) and \(T_v\) should be included in different blocks, and \(T_{A2}\) has no such constraint. If \(T_{A1}\), \(T_v\) and \(T_{A2}\) are in the same block, we classify them as single-chain sandwich attacks.
    \item \textbf{Timestamp:} The timestamp of transaction \(T_{A1}\) must fall between those of the start transaction on source chain \(T_s\) and end transaction on destination chain \(T_v\). The timestamp of transaction \(T_{A2}\) should be controlled within a suitable range, where within \(num\) block times after \(T_v\).

\end{itemize}

\subsection{Detection Settings}
We select Symbiosis~\cite{SymbiosisDocumentation} and deBridge~\cite{deBridge} as our cross-chain transaction data sources. At the time of writing, Symbiosis has processed over 4,930,238 cross-chain transactions, with a cumulative volume exceeding 9.7 billion USD since November 2022~\cite{SymbiosisExplorer}, while deBridge has processed more than 10,432,492 transactions, reaching a total volume of 23.6 billion USD since December 2022~\cite{deBridge}.  
Both protocols follow a similar design pattern: when a user submits a cross-chain transaction, the bridge contract on the source chain emits an event that contains the calldata intended for execution on the destination chain. Concretely, Symbiosis emits an OracleRequest event\footnote{0x5523985926Aa12BA58DC5Ad00DDca99678D7227E on Ethereum}, whereas deBridge emits a CreatedOrder event\footnote{0xeF4fB24aD0916217251F553c0596F8Edc630EB66 on Ethereum}, each carrying the full execution payload.
An attacker can locally reconstruct the bridge contract’s function logic on the destination chain and iteratively parse that calldata to extract the information they need, for example, which destination chain and liquidity pool is targeted, how much token X will be swapped, and the minimum acceptable amount of token Y. With this information, the attacker can then perform a cross-chain sandwich attack.

The Symbiosis protocol relies on external DEX aggregators to compute the optimal swap path, so the calldata emitted in the event contains the aggregator’s swap execution logic. Specifically, taking 1inch~\cite{1inch} as an example, Symbiosis ultimately calls the \(swap\) function in \(AggregationRouterV5\) to perform the token exchange in the corresponding pools. In practice, however, we found that the \(executor\) contract\footnote{0x5141B82f5fFDa4c6fE1E372978F1C5427640a190 on Ethereum} responsible for parsing the calldata in \(swap\) is not open source, so the exact format of the calldata is unknown to us. To address this, we use Tenderly~\cite{Tenderly} to replay the transactions corresponding to the calldata on specific historical blocks and record the swap details along the contract execution path.

\begin{figure}[htbp]
    \centering
    \subfigure[Symbiosis]{\includegraphics[width=.47\linewidth]{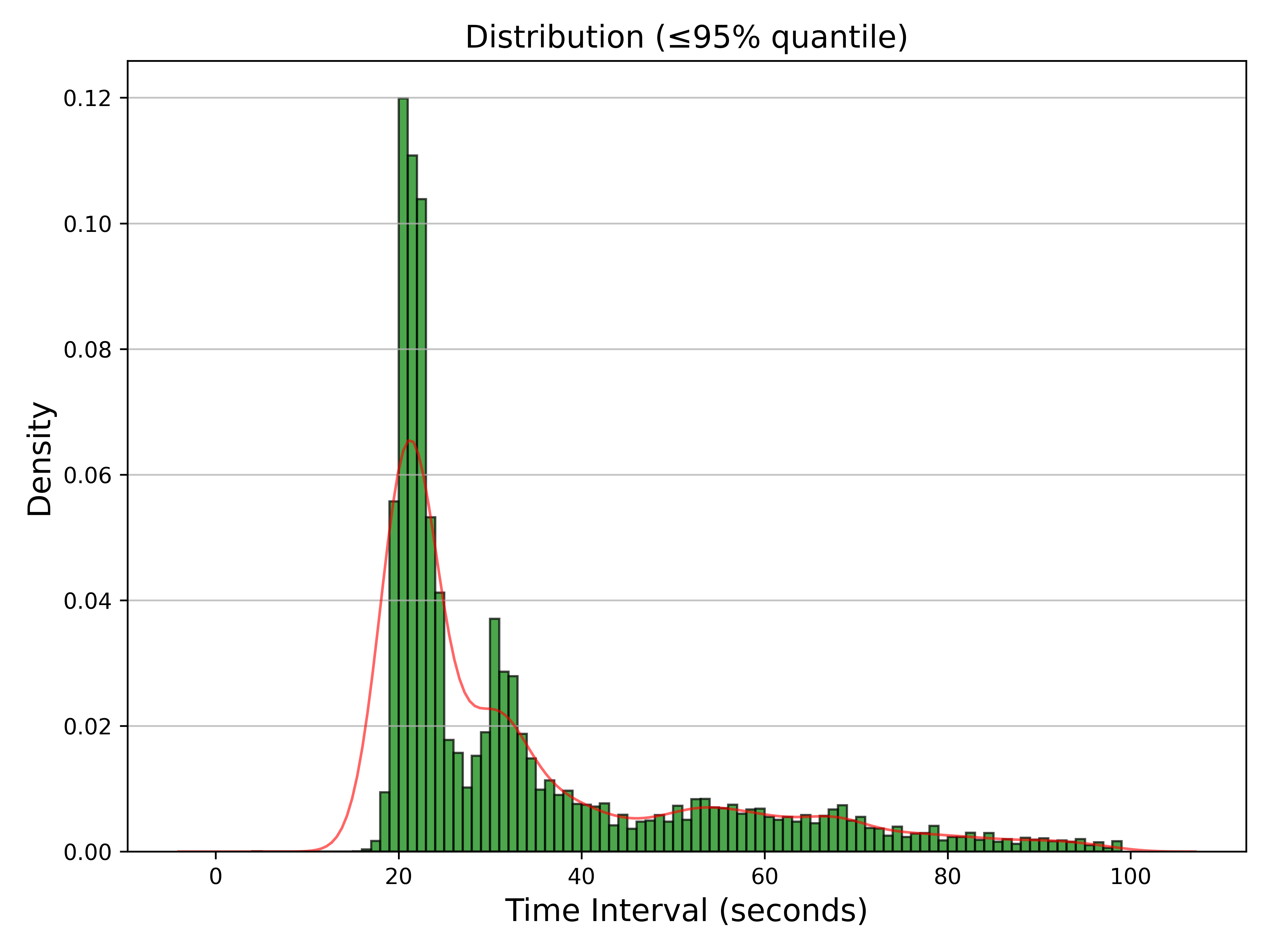}}
    \hfill
    \subfigure[deBridge]{\includegraphics[width=.47\linewidth]{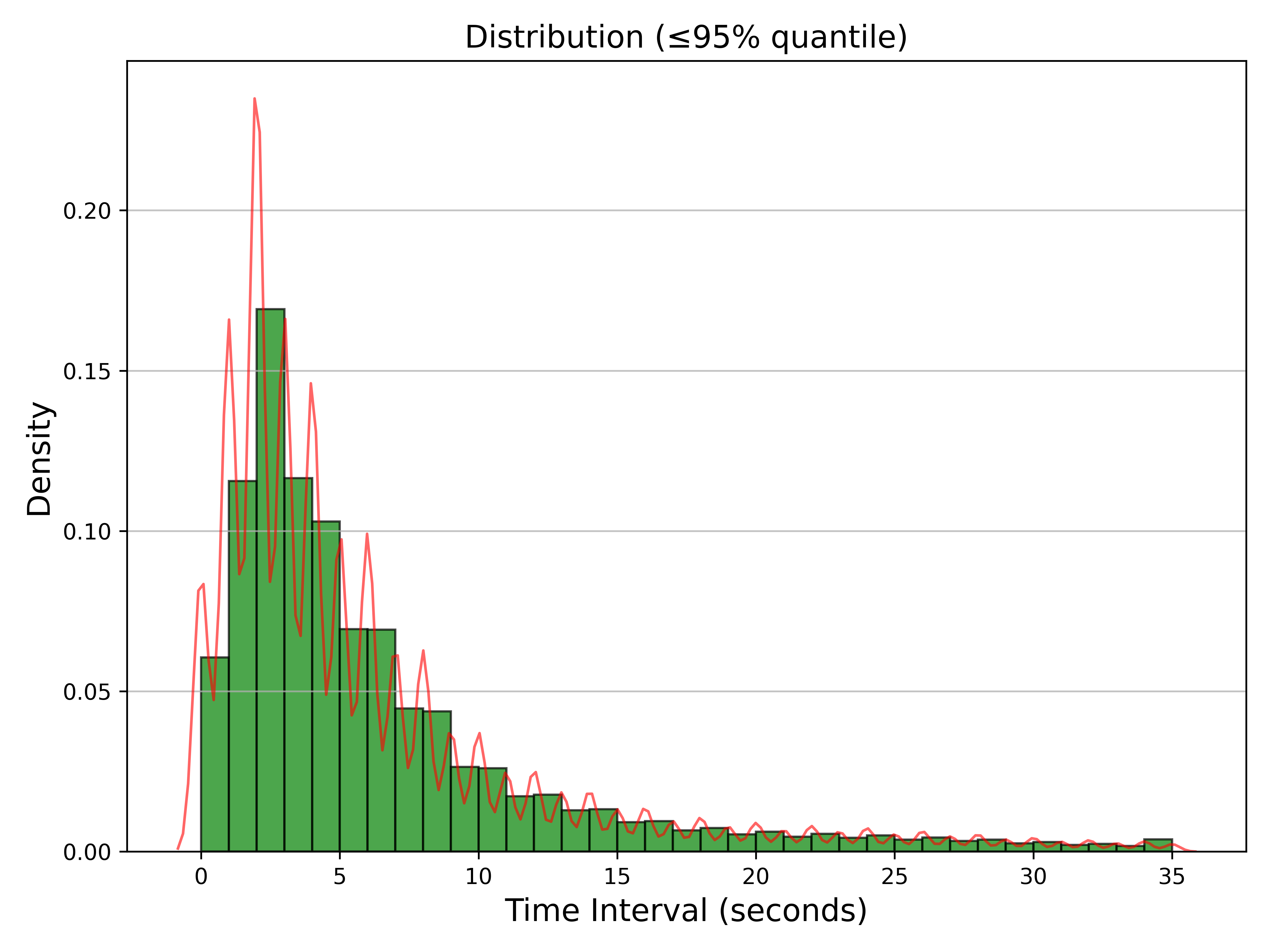}}
    \caption{{The distribution of the time interval between the event is emitted on the source chain and the swap is executed on the destination chain.}}
    \label{time}
\end{figure}

\begin{table*}[!htbp]
\caption{\textbf{Top 10 blockchain pairs by profit}}
\label{chainpair}
\centering
\scalebox{0.87}{
\begin{tabular}{lllllllllll}
\toprule
Source chain&Base&Arbitrum&Sonic&Ethereum&Polygon&Base&Ethereum&Avalanche&Avalanche&BSC\\
Destination chain&Ethereum&Ethereum&Ethereum&BSC&Ethereum&BSC&Base&BSC&Ethereum&Ethereum \\
Profit(USD)&74530&57396&50669&45752&34857&33361&28841&24571&24290&24232 \\
Percentage&13.35\%&10.28\%&9.08\%&8.20\%&6.25\%&5.98\%&5.17\%&4.40\%&4.35\%&4.34\% \\
\bottomrule
\end{tabular}
}
\end{table*}

We collect historical cross-chain transaction data from Symbiosis (November 21, 2022 to April 4, 2026) and deBridge (December 21, 2022 to April 4, 2026)\footnote{\href{https://www.kaggle.com/datasets/anonymous8397458/cross-chain-sandwich-datasets}{Cross-chain Sandwich Dataset}}. We first perform an initial filtering of the dataset based on two criteria.
The first criterion is that the swap on the destination chain must occur within a liquidity pool and must not involve stablecoin-to-stablecoin exchanges.
The second criterion requires that the time interval between \(T_s\) and \(T_v\) falls within a reasonable range. In practice, we observe that this interval can exceed 10 minutes in both protocols; however, such delayed cross-chain transactions provide limited analytical value for our study. As illustrated in Figure~\ref{time}, the interval distribution shows that 95\% of transactions are completed within 100 seconds in Symbiosis and within 35 seconds in deBridge. Accordingly, we adopt 100 seconds and 35 seconds as the maximum allowable intervals for the two protocols, respectively.
We then apply the detection heuristics described above to find instances of such attacks within the filtered transactions.

\subsection{Dataset}

To support reproducibility, we release both the normalized source cross-chain records and the corresponding sandwich-analysis records. The Symbiosis dataset was collected through the API used by its public explorer and covers completed cross-chain transactions from November 21, 2022 to April 4, 2026. For deBridge, we contacted the protocol team and obtained access to its historical-data API, through which we queried the available cross-chain orders from December 21, 2022 to April 4, 2026.

The released source datasets represent completed cross-chain transactions as pairs \((T_s,T_v)\), where \(T_s\) is the source-chain transaction and \(T_v\) is the corresponding destination-chain transaction. Symbiosis API records provide the core transaction-pair information together with the source and destination execution times, while deBridge API records additionally include order identifiers, chain identifiers, token addresses and symbols, decimals, and raw or final transfer amounts. The released sandwich-analysis records further attach detector evidence to each cross-chain transaction, including the victim transaction, the attacked pool, candidate swap transactions before and after the victim, and the matched sandwich triples \((T_{A1},T_v,T_{A2})\).

\subsection{Empirical Results}
\label{result}

\noindent\textbf{Transactions.}
In total, we collect 4,765,366 cross-chain transactions from the two protocols. By applying the filtering criteria\footnote{Cross-chain transactions in deBridge more often result in direct transfers on the destination chain; a swap is only performed when the solver’s wallet does not hold the required token.} and Algorithm~\ref{detect_algo} to these transactions, we identify 19390 potential cross-chain sandwich attack pairs. This accounts for only 0.4\% of the total, as this type of attack remains largely underexplored. Among them, only 283 pairs occur within the same block, which represents single-chain sandwich attacks, they account for only 1.4\% of the total. Table~\ref{pooltims} provides an overview of pools that were attacked more than 100 times in the dataset. We observe that the StableCoin\(\leftrightarrow\)WETH pools are the most frequently targeted, accounting for 37.1\% of all identified attacks.

\begin{figure}[htbp]
    \centering
    \includegraphics[width=\linewidth]{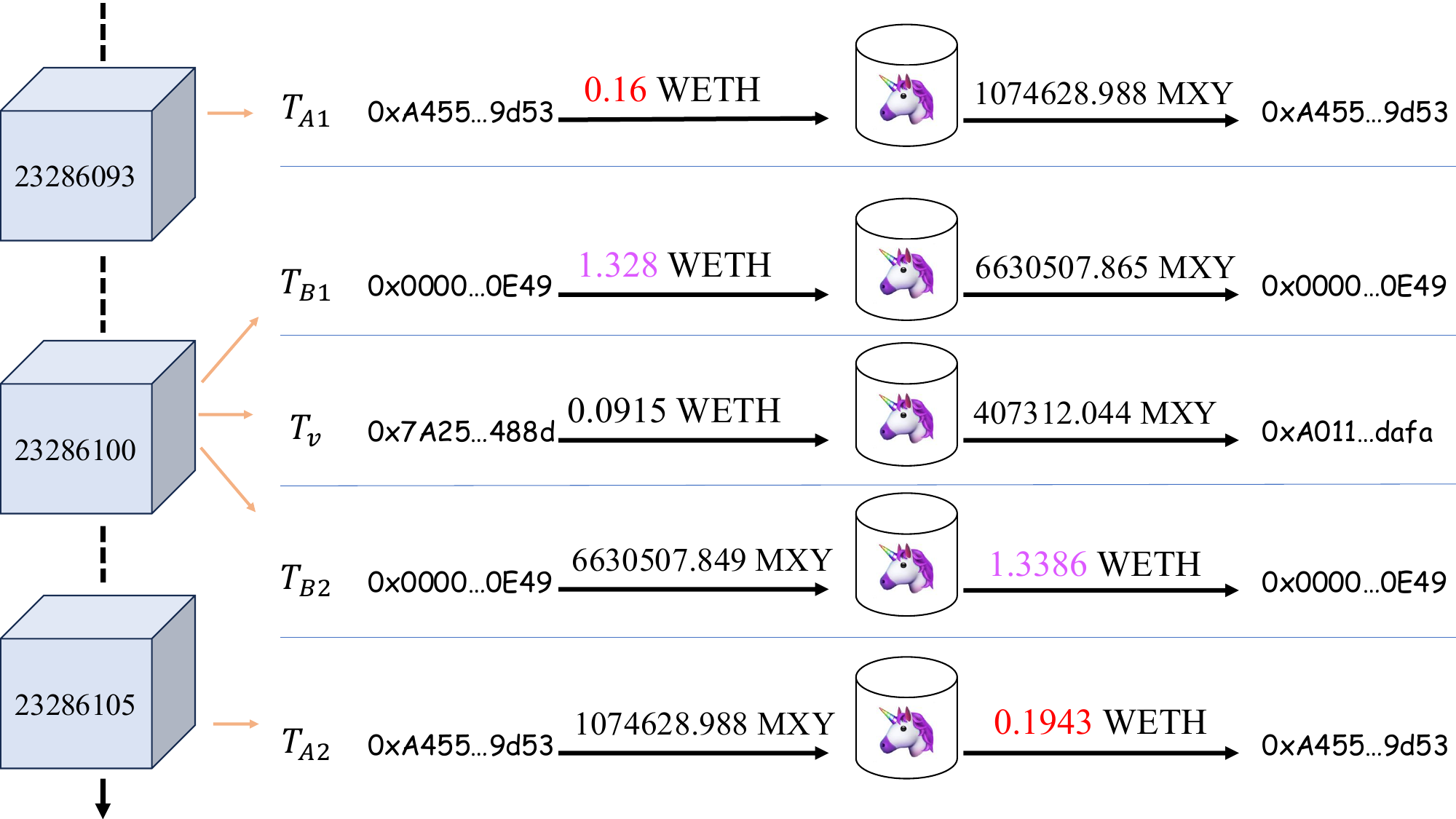}
    \caption{Case study of one cross-chain sandwich attack on the Ethereum Mainnet}
    \label{example}
\end{figure}

We provide a perfect case of a cross-chain sandwich attack in which both single-chain and cross-chain sandwich attacks coexist on Ethereum Mainnet: transactions (\href{https://basescan.org/tx/0x0aaf8f5aac0b43bfcba7e7f227028c6b3d52ade3fba6ab056ecb43ef0919b44f}{\textcolor{cyan}{0x0aaf...b44f}},
\href{https://etherscan.io/tx/0x9d5f874f8fde164a45cc758611424c6e91e60ed98f265f1a6ed1f11b5482d548}{\textcolor{cyan}{0x9d5f...d548}}, \href{https://etherscan.io/tx/0xef068c4ae7e56a54a72432887dc9faa6ea2f7ce9606d8ddc758c975970692d7a}{\textcolor{cyan}{0xef06...2d7a}}, \href{https://etherscan.io/tx/0xe16b2360053139b1d2991c26380529007ddd20ca6bb8d5ec59076bb4853282c0}{\textcolor{cyan}{0xe16b...82c0}}, \href{https://etherscan.io/tx/0xc07da1dea2dba2baf985e2f774342f3ebdec551d40905bf9400646e04f612b8f}{\textcolor{cyan}{0xc07d...2b8f}}, \href{https://etherscan.io/tx/0xe2a1c948e86bdc4785bf3fee56218a16d823ccbd56662c0d1d490249a94986de}{\textcolor{cyan}{0xe2a1...86de}}) correspond to components (\(T_s\), \(T_{A1}\), \(T_{B1}\), \(T_v\), \(T_{B2}\), \(T_{A2}\)) described above, and this example is shown in Figure~\ref{example}.
\(T_v\) attempted to swap WETH for MXY. \(\mathcal{A}\) placed \(T_{A1}\) at block 23286093, then \(T_v\) was successfully executed at block 23286100. Concurrently, \(T_v\) was also targeted by \(\mathcal{B}\), and \(\mathcal{B}\) is labeled as a MEV Bot in this case. However, existing MEV monitoring platforms, such as EigenPhi\footnote{https://eigenphi.io/mev/ethereum/tx/\\0x4c0d2d55daef8fce10b340445a14789532e47a9c0d45bf3265489abf87b05731/}, only detect \(T_{B1}\) and \(T_{B2}\) for \(T_v\). Subsequently, \(\mathcal{A}\) placed \(T_{A2}\) at block 23286105 to extract the value. In the end, excluding gas costs, \(\mathcal{A}\) gained an additional 0.034 WETH and achieved a 21.4\% profit rate (\((\Delta x_{A2}-\Delta x_{A1})/{\Delta x_{A1}}\)), whereas the profit rate of \(\mathcal{B}\) is 0.8\%.

\begin{figure}[t]
\centering
\subfigure[Distribution of the profit in USD (\(\Delta x_{A2}-\Delta x_{A1}\))]
{
    \begin{minipage}[b]{\linewidth}
        \centering
        \includegraphics[width=\linewidth]{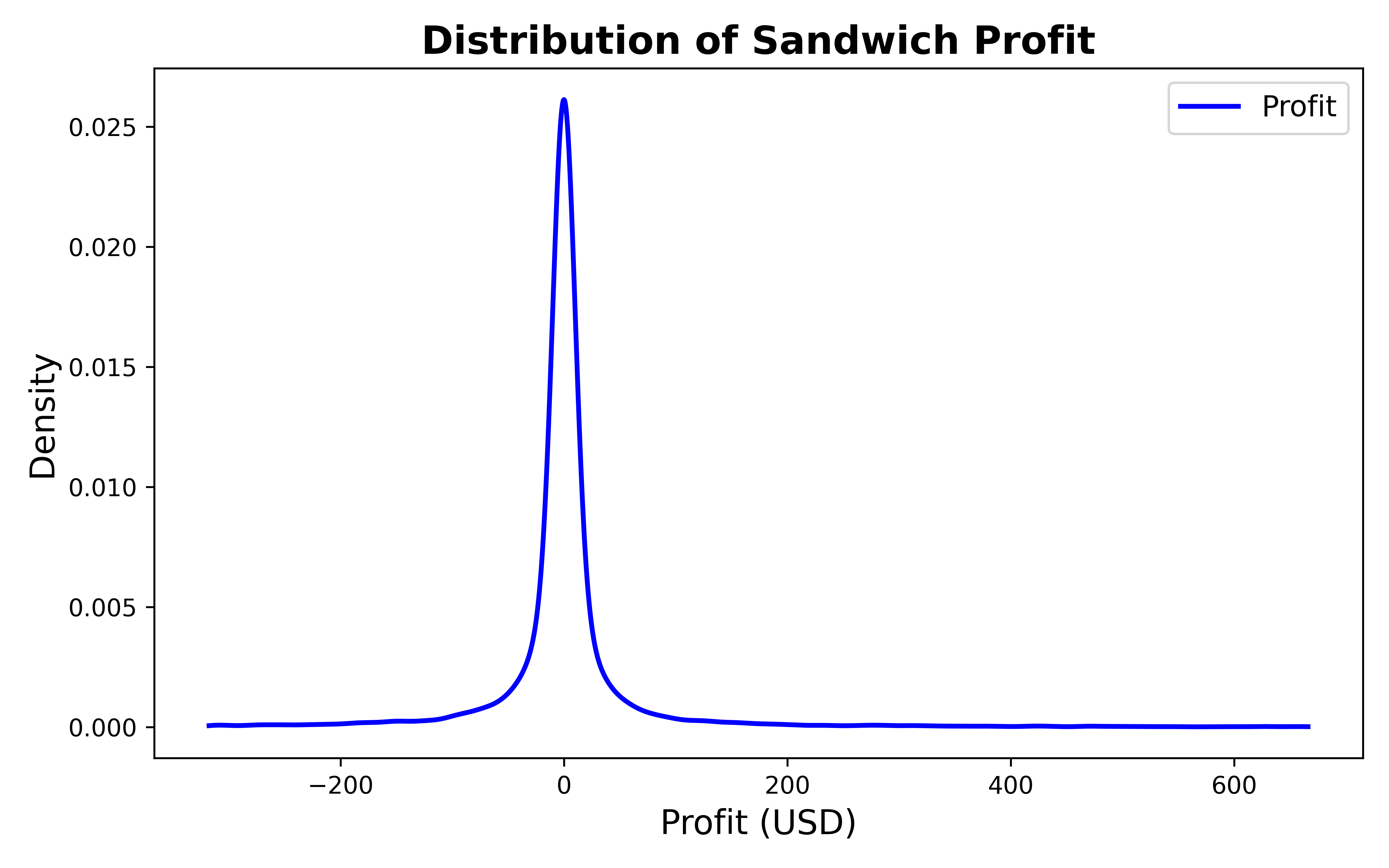}
    \end{minipage}
}
\newline
\subfigure[Distribution of the profit rate (\((\Delta x_{A2}-\Delta x_{A1})/\Delta x_{A1}\))]
{
 	\begin{minipage}[b]{\linewidth}
        \centering
        \includegraphics[width=\linewidth]{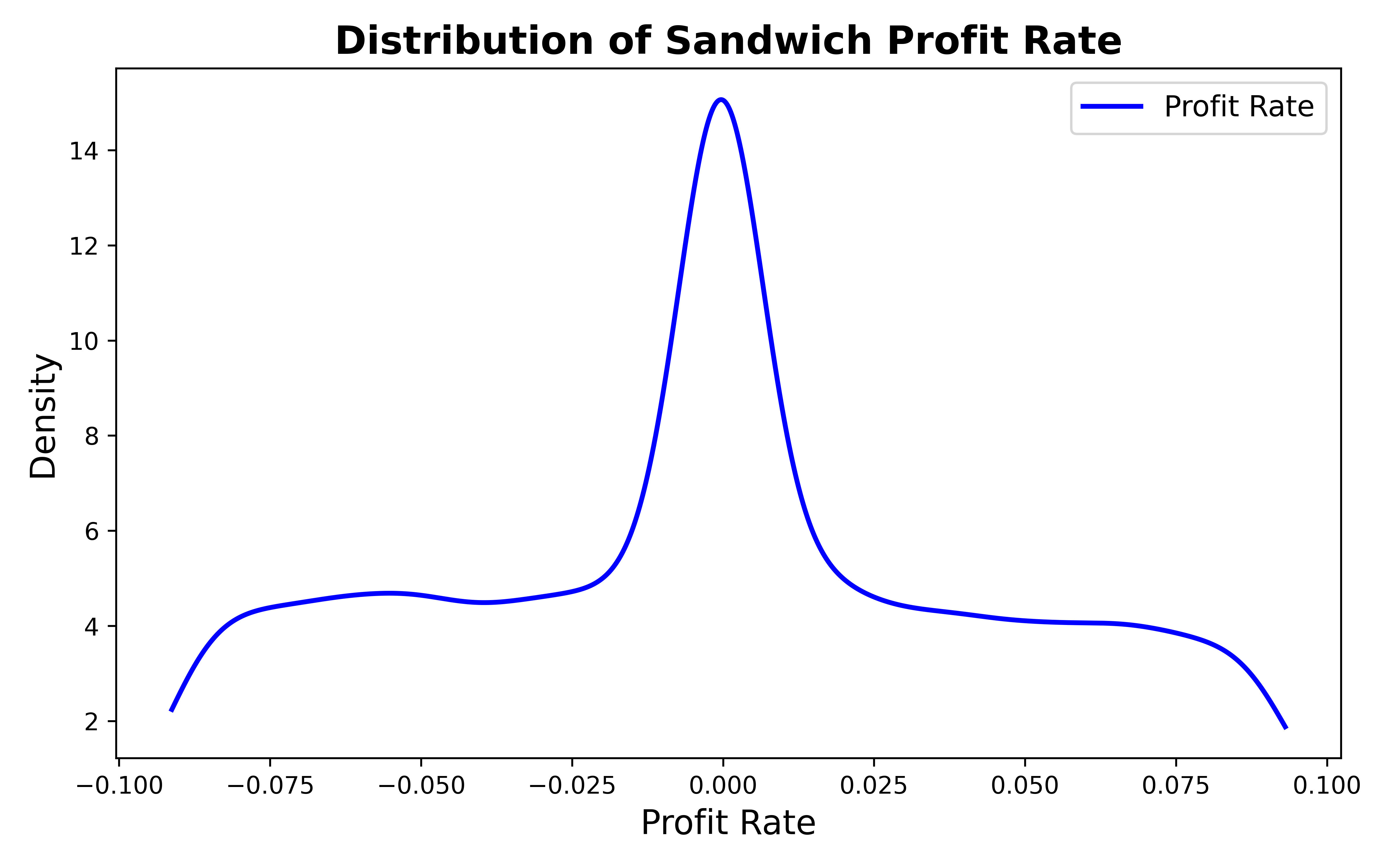}
    \end{minipage}
}
\caption{The distribution of the absolute profit and the profit rate}
\label{profit}
\end{figure}

\begin{figure*}[t]
\centering
\subfigure[Transaction relative positions for all sandwich attacks]
{
    \begin{minipage}[b]{.48\linewidth}
        \centering
        \includegraphics[width=\linewidth]{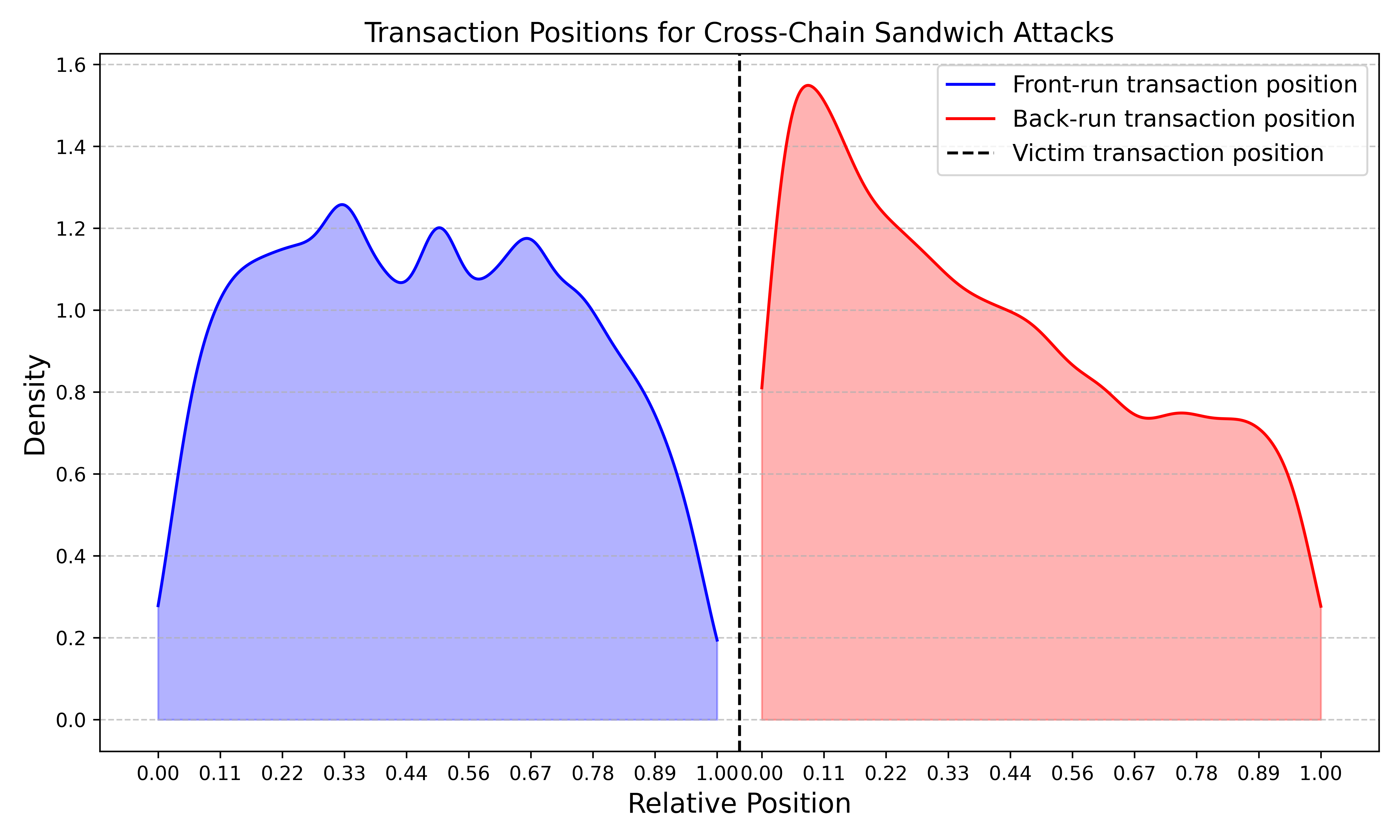}
    \end{minipage}
}
\subfigure[Transaction relative positions for profitable sandwich attacks]
{
 	\begin{minipage}[b]{.48\linewidth}
        \centering
        \includegraphics[width=\linewidth]{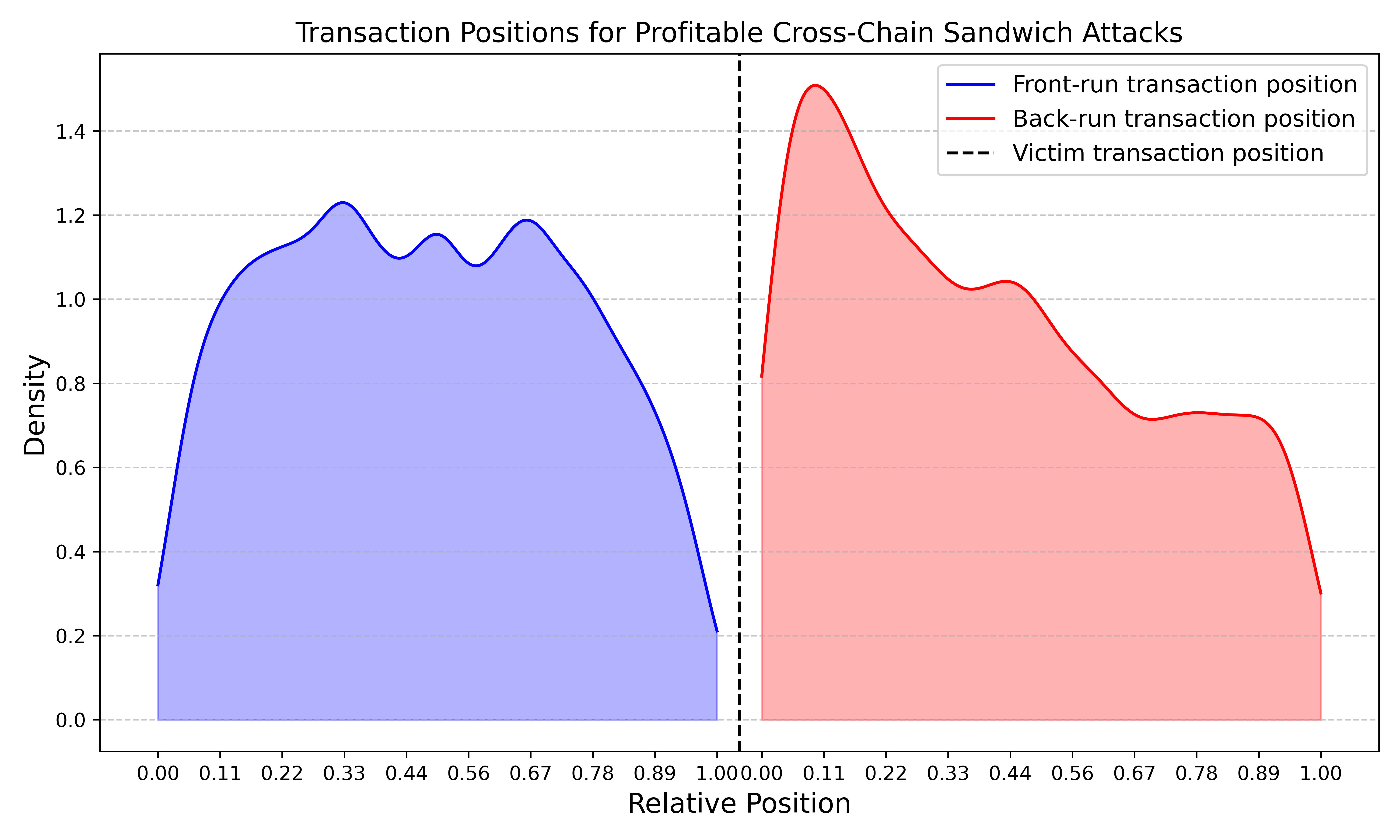}
    \end{minipage}
}
\caption{Relative positions of cross-chain sandwich attacks}
\label{positions}
\end{figure*}

\noindent\textbf{Extracted profit.}
The profit is calculated based on the token price in USD at each transaction timestamp. Overall, the total profit we calculated from our dataset with Equation~\ref{profitequ} is 558,069 USD, which is 1.75\% of a total trading volume of 31,902,228 USD, and the largest profit from a single sandwich attack is 37,103 USD. Additionally, there remains 180,340 USD worth of unexploited potential profit.

Figure~\ref{profit} presents the distributions of absolute profit and profit rate. The profit rate appears roughly symmetric around zero, while the absolute profit is predominantly positive, indicating that large-volume opportunities overwhelmingly produce positive returns.

We count the profits for each cross-chain pair and present the top 10 pairs with the highest total profits in Table~\ref{chainpair}. Among them, the chain pair Base\(\rightarrow\)Ethereum yields the highest attacker profit, with a total of 74,530 USD extracted. Figure~\ref{sankey} illustrates the proportions of profits generated on the destination chain across different cross-chain pairs. We find that cross-chain sandwich attacks are predominantly concentrated on cross-chain transactions targeting Ethereum, BSC, and Base as destination chains.

We also observed that USDC\(\leftrightarrow\)WETH is the most sandwich attack-prone token pair on the destination chain with a profit of 350,674 USD, this token pair’s pools are also the most frequently attacked (cf. Table~\ref{pooltims}). The profit accounts for 62.84\% of the total attacked volume.

In our dataset, single-chain sandwich attacks produced only 6,476 USD in profit, approximately 1.16\% of the total profits. Taken together with the single-chain sandwich transaction counts reported above, this suggests that when cross-chain sandwich attacks are already in play, little to no exploitable volume remains when the victim transaction appears in the mempool.

\noindent\textbf{Sandwich transaction positions.}
Unlike single-chain sandwich attacks that seek to place transactions immediately adjacent to the victim transaction, cross-chain sandwich attacks prioritize maximizing use of early-obtained information while avoiding direct competition with existing MEV bots. We analyze the relative block positions of transactions in each cross-chain sandwich pair, specifically, the position of \(N_{A1}\) relative to \(N_s\) and \(N_v\), and the position of \(N_{A2}\) relative to \(N_v\) and \(N_v+num\). In our experiments we set the parameter \(num=100\) in Algorithm~\ref{detect_algo}. The resulting position distribution is plotted in Figure~\ref{positions}, where the left subfigure depicts the distribution for all detected attacks, while the right subfigure shows the distribution restricted to profitable attacks. 

Empirically, we observe that \(\mathcal{A}\) tends to send \(T_{A1}\) at a point relatively close to the middle of the interval between \(T_s\) and \(T_v\). One possible explanation is that \(\mathcal{A}\) may estimate the expected duration of cross-chain transactions for a given protocol and avoid sending \(T_{A1}\) immediately after observing \(T_s\). Instead, they may wait for an appropriate time window to reduce their exposure to noisy transactions occurring between \(T_s\) and \(T_v\). In particular, delaying the attack may help avoid unfavorable pool-state changes caused by adverse or reverse swap transactions, thereby reducing the risk of incurring losses. In contrast, \(\mathcal{A}\) tends to submit \(T_{A2}\) immediately after \(T_v\) is executed in order to realize profit. We observe that the distribution of relative positions for all sandwich attacks closely resembles that of profitable sandwich attacks, indicating that the attackers follow similar strategies. However, their specific profits are influenced by the impact of intermediate noisy transactions.

\noindent\textbf{Sandwich transaction gas price.}
For each sandwiching transaction pair, we examine the relationship between the gas price of \(T_{A1}\) and that of \(T_v\), as well as between the gas price of \(T_{A2}\) and \(T_v\). The difference of gas prices is shown in Table \ref{gasdifference}. 
For \(T_{A1}\), we find that in 63.41\% of cross-chain sandwich attacks, the gas price of $T_{A1}$ is only 0--1 Gwei higher than that of the victim transaction $T_v$. This gas-price relationship is similar to that observed in conventional single-chain sandwich attacks. However, unlike single-chain attacks, $T_{A1}$ and $T_v$ are included in different blocks in the cross-chain setting and therefore do not directly compete for transaction ordering within the same block.

One possible explanation is that attackers may use the network-recommended gas price as a reference when submitting $T_{A1}$. In particular, \(\mathcal{A}\) may query \texttt{eth\_gasPrice} and set the gas price slightly above the recommended value to increase the likelihood of timely inclusion. Because such cross-chain transactions are generally completed within a short time window, the network-recommended gas price may remain relatively stable between the submission of $T_{A1}$ and $T_v$. Meanwhile, the relayer usually uses the network-recommended gas price value to send \(T_v\), so the two transactions are expected to have similar gas prices.

We observe that in 64.34\% of cross-chain sandwich attacks, the gas price of $T_{A2}$ is lower than that of $T_V$. This observation is intuitive, as \(\mathcal{A}\) intends to execute $T_{A2}$ after $T_V$ in order to capture the resulting profit.

\begin{table}[htbp]
\caption{\textbf{The differences of gas prices}}
\label{gasdifference}
\centering
\scalebox{0.87}{
\begin{tabular}{llrr}
\toprule
Difference & Range & Number & Percentage\\
\midrule
\multicolumn{4}{l}{\(\delta_1=GasPriceT_{A1}-GasPriceT_v\)}\\
\(\delta_1\) & \(\delta_1<0\) GWei
    & 5906 & 30.46\%\\
\(\delta_1\) & \(0\ \mathrm{GWei}\leq\delta_1<1\ \mathrm{GWei}\)
    & 12295 & 63.41\%\\
\(\delta_1\) & \(1\ \mathrm{GWei}\leq\delta_1<10\ \mathrm{GWei}\)
    & 875 & 4.51\%\\
\(\delta_1\) & \(10\ \mathrm{GWei}\leq\delta_1\)
    & 314 & 1.62\%\\
\(\delta_1\) & Total
    & 19390 & 100\%\\
\midrule
\multicolumn{4}{l}{\(\delta_2=GasPriceT_v-GasPriceT_{A2}\)}\\
\(\delta_2\) & \(\delta_2<0\) GWei
    & 12475 & 64.34\%\\
\(\delta_2\) & \(0\ \mathrm{GWei}\leq\delta_2<1\ \mathrm{GWei}\)
    & 6473 & 33.38\%\\
\(\delta_2\) & \(1\ \mathrm{GWei}\leq\delta_2<10\ \mathrm{GWei}\)
    & 427 & 2.20\%\\
\(\delta_2\) & \(10\ \mathrm{GWei}\leq\delta_2\)
    & 15 & 0.08\%\\
\(\delta_2\) & Total
    & 19390 & 100\%\\
\bottomrule
\end{tabular}
}
\end{table}

\noindent\textbf{Attack addresses.}
From the 19,390 detected attacks, we identified a total of 3,155 attacking addresses, including 2,241 EOA addresses and 914 contract addresses. Among these addresses, the top 200 attackers were responsible for 13,941 attacks, meaning that just 6.3\% of the attackers accounted for 71.9\% of all detected attacks. The table\ref{eoaaddress} shows the top 10 EOA addresses ($T_f.from = T_b.from$) by attack frequency. We find that these EOA addresses are predominantly concentrated on the Base chain, and their transaction histories contain a large proportion of StableCoin-to-ETH swaps.

\begin{figure}[!htbp]
    \centering
    \includegraphics[width=\linewidth]{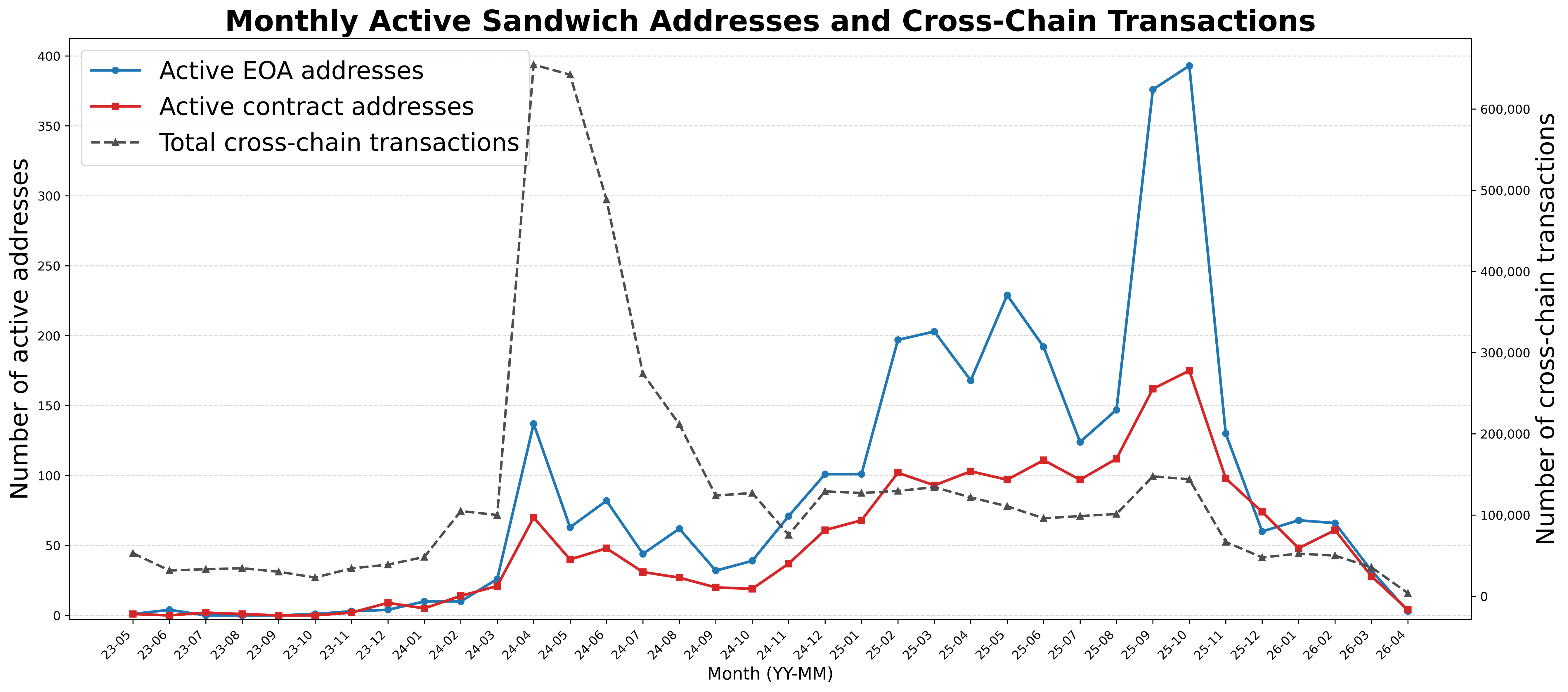}
    \caption{Number of monthly active attack EOA and smart contract addresses.}
    \label{monthly}
\end{figure}

Figure~\ref{monthly} illustrates the number of active attacker addresses per month\footnote{An address is considered active in a given month if it performed at least one attack during that month} and the total number of cross-chain transactions processed by deBridge and Symbiosis each month. We observe a significant surge in cross-chain sandwich attacks in September 2025. However, this increase does not exhibit an apparent correlation with the overall number of cross-chain transactions.

\noindent\textbf{Empirical parameters.}
We computed the profit rates for all detected cross-chain sandwich attacks. We observed several extreme cases in which positive returns approached or exceeded 1 and negative returns approached -1. For example, cross-chain attack transaction pairs (
\href{https://etherscan.io/tx/0x970d66a5e22f83a79c3725c037c528647d777452ca2fb1358dc81667801e6948}{\textcolor{cyan}{0x970d...6948}},
\href{https://etherscan.io/tx/0xc94e5177dfe52e40077bdd33af44193c8b957d3ffe1ae79567fb180e90f29f4a}{\textcolor{cyan}{0xc94e...9f4a}},
\href{https://etherscan.io/tx/0xc12ee93d02d1c92ec498c19fa472fb9de193296bb3ef4c6c12b9580ec1e75fe5}{\textcolor{cyan}{0xc12e...5fe5}},
) and (
\href{https://etherscan.io/tx/0x14c1df067a5eb316b0631ce77be82c047018883ceaee8ae30b133b3542a0de16}{\textcolor{cyan}{0x14c1...de16}},
\href{https://etherscan.io/tx/0xd1f2a441bded8d9cbd3e6dd5bfd7f6aa7433a67d0aeef3f7fa7ea9b8e6a9c35c}{\textcolor{cyan}{0xd1f2...c35c}},
\href{https://etherscan.io/tx/0x9b2ada7afc92c5b0218202492041dea59bae741d987f5d20ef4913afb35d31b1}{\textcolor{cyan}{0x9b2a...31b1}},
)
had profit rates of \(0.62\) and \(-0.72\) respectively. These transactions all involve swaps containing altcoins, whose prices are highly susceptible to manipulation and often experience large price fluctuations. Therefore, such transactions provide limited value for our sandwich attack analysis. We filter extreme returns using the respective 95th-percentile thresholds of positive and absolute negative return rates, retaining 18,422 attack pairs. The conditional means of positive and negative returns in this retained sample are \(r^+=0.0373\) and \(r^-=-0.0384\), respectively.

We estimate \(p\) and \(q\) on the same retained sample. For each attack pair, we extract \(T_1...T_{n}\) that occurred between the timestamps of \(T_s\) and \(T_v\), and examine their impact on the liquidity pool's state. In 2,525 attack pairs, no transactions occurred between these timestamps, yielding \(q=13.71\%\). Among the remaining 15,897 attack pairs where \(T_1...T_{n}\) existed, 11,766 attack pairs exhibit final states in which \(T_{A1}\) did not incur loss after the execution of \(T_v\), resulting in an empirical value of \(p=74.01\%\). Thus, \(q+(1-q)p=77.58\%\). Using these parameters from the same sample, the calibrated model yields
a model-implied profit rate estimate of
\begin{equation*}
    (q+(1-q)p)r^+ + (1-q)(1-p)r^-\approx2.03\%.
\end{equation*}

\begin{table}[t]
\caption{\textbf{Top 10 EOA addresses by numbers of attacks}}
\label{eoaaddress}
\centering
\scalebox{0.8}{
\begin{tabular}{lll}
\toprule
Address&Chain&Attacks\\
\midrule
0xc8e42017d04103377feecc676de30164476d23a1&BSC&216\\
0xc1ecfe1d7f51674c6e7baf421bf9cf14e75f6d43&Base&185\\
0xd0adade7cf2b6dd8b2f673c1b699dd21062a15ea&Base&181\\
0x7fd624f3f97a7dd36195e8379f28db6147c270ff&Base&138\\
0x3a53cc8a26691fd434f28dea2ce9a33d85730997&Base&130\\
0x5475ec93b04cc3babfbf4eddabb9b84cb5ba149d&Base&122\\
0xf0eece173ff0a8873769ee0f2a6fdffa0da37b1c&BSC&119\\
0xa1c9e0381150eb24d164ef0db18e606e65f751af&Base&97\\
0xf6cd5fb740bc1f4b21989d10b5e4366089f6fbe1&Base&92\\
0xec440921f3a4e00b5962a8c6fe68229d3a7785ad&BSC&90\\

\bottomrule
\end{tabular}
}
\end{table}

\noindent\textbf{Private transactions.}
To complement our analysis, we obtained Ethereum mempool data from mempool.guru~\cite{mempool.guru,yang2024sok} covering the period of our cross-chain dataset. These data include first-seen timestamps, which record when a monitored node first observes a transaction, and a set of transactions classified as private. We use these data to examine all 1,461 candidate cross-chain sandwich triples targeting Ethereum.

Notably, 1,163 triples (79.60\%) have \(T_v\) classified as private. This is consistent with our expectation that destination-chain privacy does not address source-chain information leakage: publicly emitted events can reveal the intended swap even when its execution transaction is submitted privately. Such submission may also reduce competition from conventional MEV bots that rely on the destination-chain public mempool, potentially leaving more opportunities for cross-chain attackers who learn about the swap from source-chain events. We also identify 1,197 triples (81.93\%) whose \(T_{A1}\) is classified as private, consistent with attackers using private submission to protect their own front-running transactions from being sandwiched.

Finally, only 25 triples have first-seen records for both \(T_{A1}\) and \(T_v\) at a common monitor in the mempool.guru dataset. Of these, 20 satisfy \(\operatorname{firstseen}(T_{A1})<\operatorname{firstseen}(T_v)\) at every common monitor, consistent with the temporal ordering expected for cross-chain sandwich attacks.

\noindent\textbf{Limitations.}
Currently, we have only collected cross-chain transaction data from EVM-based chains, we parsed them with event and function shown in Table~\ref{events}. However, in the Symbiosis protocol, there also exist destination chains of the UTXO type, such as Bitcoin. Since these chains lack smart contract support, we are currently unable to access the exact data format used by the protocol in such cases, and therefore can not parse the corresponding transactions. As a result, the potential profits from these cross-chain interactions remain unknown and the real profit is larger than that we calculated.

In addition to Symbiosis and deBridge, other cross-chain protocols, such as Thorswap~\cite{ThorSwap}, also meet the prerequisites for the cross-chain sandwich attack. Thorswap reached a peak monthly swap volume of 11.18 billion USD~\cite{ThorStats}. It maintains its own liquidity pools and executes swaps via its relay chain. However, it does not provide an interface to query pool swap information on the relay chain, preventing us from analyzing its historical data. By examining more protocols of this type in the future, we will be able to obtain a more accurate assessment of cross-chain sandwich attack potential. 

Moreover, the occurrence of two transactions immediately before and after a potential victim transaction does not always indicate an actual cross-chain sandwich attack, as such a pattern may also arise coincidentally. Therefore, our heuristic-based detection approach may inevitably classify some legitimate transactions as attacks, leading to false positives. Since there is no publicly available ground truth for validation, our reported results should be interpreted as approximate measurements rather than exact attack counts.

\subsection{Discussion of Future Strategy}
The foregoing analysis is based on historical data. However, from the attacker’s perspective there exist multiple viable strategies in the cross-chain setting. The first strategy corresponds to the one described in Section~\ref{attacker model}: upon observing a profitable cross-chain event, the attacker immediately submits a front-running transaction. After the victim transaction is executed on the destination chain, it immediately submits the back-running transaction and has no other intermediate actions. This strategy is based on the attacker’s belief in the empirical value of probability \(q\). 

A second class of strategies involves dynamic decision-making: the attacker adaptively adjusts behavior in response to noisy transactions. For example, if the reverse-direction transaction volume exceeds the attacker’s tolerance, the attacker may cut losses early. Conversely, when same-direction noisy transactions appear, some attackers may adopt a greedy policy by submitting a small back-running transaction to capture a portion of the profit. In extreme cases, a large and same-direction transaction pushes the victim transaction’s execution beyond its slippage tolerance and causes the transaction revert, the attacker may place the back-running transaction immediately after that large transaction to realize even greater profit. However, because the Symbiosis protocol does not expose information about cross-chain transactions that are reverted, we can only discuss this possibility qualitatively and can not identify empirical instances to validate the existence of such strategies.

It is worth noting that some cross-chain protocols do not publish their event data formats (i.e., the corresponding smart contract code is not open source). In such cases, an attacker must locally simulate the calldata execution on the most recent block to infer the swap path on the destination chain and then place \(T_{A1}\) in the relevant pool. This approach introduces uncertainty because the actual execution time of \(T_v\) is unknown. If execution is delayed, the state of blockchain may change and the calldata’s execution result may different. That can cause the attacker to misjudge the details of \(T_v\) and thus increase the risk that the attacker will incur a loss.

\section{Mitigation}
The information necessary for cross-chain sandwich attacks is obtained on the source chain. Consequently, future changes to mempool opacity~\cite{weintraub2022flash,choudhuri2024mempool,choudhuri2025practical} or to transaction-ordering mechanisms~\cite{kelkar2020order,kelkar2022order} on the destination chain would not prevent this attack. 

To eliminate cross-chain sandwich attacks in such protocols, one must either prevent information emitted on the source chain from being publicly exposed, or ensure that any publicly available information is useless to an attacker (cut \ding{205} or \ding{206} in Figure~\ref{framework}). This argument relies on the assumption that the protocol itself is honest and reliable. If the protocol designers or operators are malicious, the attack can not be stopped. In the worst case, a malicious designer can read cross-chain information directly from the frontend or other privileged channels and execute such sandwich attacks. They observe the information (\ding{172} in Figure~\ref{framework}) strictly earlier than any external attacker (\ding{205} in Figure~\ref{framework}). Based on this assumption, we present several potential countermeasures against cross-chain sandwich attacks.

\noindent\textbf{Private relayers.} A cross-chain protocol could choose not to publish events on-chain, and instead send the corresponding cross-chain information privately to its own relay nodes or a set of trusted relays. While this design can completely prevent cross-chain sandwich attacks, it also means the protocol must give up the market of decentralized relay network and take on the cost of maintaining its own relay network. Moreover, this increases the level of centralization in the protocol and reduces users’ trust in it.

\noindent\textbf{Encrypted message.} A protocol can encrypt the cross-chain swap information on the source chain and only reveal the plaintext on the destination chain after the relayer forwards the encrypted message. However, decrypting data directly on the bridge contract is impractical. First, all key variables and decryption logic within a smart contract are publicly visible. Second, employing cryptographic techniques such as DKG~\cite{kate2024non} to achieve decentralized key management would incur extremely high gas costs and introduce significant time delays, making it unsuitable in practice. Therefore, some entity must have access to the plaintext of the event in order to execute it. The protocol can introduce a trusted off-chain decryption committee: after receiving the encrypted event from the relayer, the committee selects a member to decrypt the event and execute the calldata logic on the destination chain.
In this design, even if a relayer or an attacker observes the event emitted on the source chain, they can not extract useful information and therefore cannot perform a cross-chain sandwich attack. 

\noindent\textbf{Use real-time swap calldata.}
The root cause of cross-chain sandwich attacks is that such cross-chain protocols compute the exact swap calldata on the source chain (via DEX aggregators like 1inch~\cite{1inch} and OpenOcean~\cite{openocean}) so the frontend can show the user a precise expected output. That calldata which contains the chosen pools and execution path is then sent to the destination chain and becomes observable to attackers.

A simple and effective fix is to move the swap calldata-generation step to the destination chain. On the source chain, the relayer should only receive the minimal necessary information, such as that the user wants to swap \(\Delta x\) token X for token Y, without including any details about the pool, path, or routing. Because aggregators’ optimal paths change over time, an attacker who only sees this rough intent can not predict the exact future calldata or chosen route. Guessing the specific pool and submitting a front-running transaction carries risk, as a rational attacker, they would not attempt such an attack.

\section{Related Works}

\noindent\textbf{Maximum extractable value.}
Eskandir et al.~\cite{eskandari2019sok} are the first to introduce the concept of front-running from traditional financial markets into DeFi. Subsequently, Flash Boys 2.0~\cite{daian2020flash} provide a study on priority gas auctions, where the winner gains priority in transaction ordering. This work introduced the concept of MEV and empirically demonstrated that MEV poses a significant threat to Ethereum. Torres et al.~\cite{torres2021frontrunner} and Qin et al.~\cite{qin2022quantifying} conducted a systematic analysis based on historical Ethereum transaction data and showed that many miners have already been extracting MEV through gas bidding. In addition, Qin et al.~\cite{qin2022quantifying} evaluate the profits miners obtained from arbitrage, liquidation, and sandwich attacks, respectively. More specifically, Zhou~\cite{zhou2021high} formally define the sandwich attack problem in AMM. They analyze it both theoretically and empirically from the attacker’s perspective, identifying the conditions under which such attacks can yield profit. 

To detect the MEV behaviors, some works explore learning-based detection of malicious transactions and MEV bots at scale~\cite{wu2025hunting,ma2025surviving,niedermayer2024detecting}. Li et al.~\cite{li2023demystifying} conduct a comprehensive study of Flashbots bundles and discovered 17 previously unknown DeFi MEV strategies by leveraging their detection tool.

There are currently several effective countermeasures against sandwich attacks. Proposer-Builder Separation (PBS) mechanism~\cite{yang2025decentralization} in Ethereum splits block production into two roles. Builders construct blocks and include transactions, while proposers only select which block to propose. This separation improves fairness in transaction ordering. Aequitas provided by Kelkar et al.~\cite{kelkar2020order} is a protocol designed to achieve order fairness at the consensus layer, and it is later extended to a permissionless setting~\cite{kelkar2022order}. Furthermore, there are some works that leverage the encrypted mempool~\cite{choudhuri2024mempool,choudhuri2025practical,bormet2025beat} to hide transaction content until ordering is finalized.

We can summarize that all existing mitigations aim to ensure fair transaction ordering within the block containing the victim transaction. However, cross-chain sandwich attacks do not rely on transaction ordering within that block, they place front-running transaction in the block preceding the target block. Therefore, all of the aforementioned methods are ineffective against this type of attack.

\noindent\textbf{Cross-domain MEV.}
Cross-domain MEV arises when extractable value spans different domains, such as Layer 1 to Layer 2 or cross-chain environments. Its essence lies in exploiting manipulation opportunities caused by network communication delays, state differences, or information asymmetry between different domains.

Cross-chain arbitrage is a common form of cross-domain MEV, arbitrageurs observe price differences of the same token across different chains and execute rapid buy and sell trades to capture profits. Oz et al.~\cite{oz2025cross} analyze one year of data across nine blockchains and identified 868.64 million USD worth of cross-chain arbitrage transactions. Additionally, they observe a centralization trend in cross-chain arbitrage, with five addresses generate more than half of these transactions. 
Gogol et al.~\cite{gogol2024cross} investigate non-atomic arbitrage in cross-rollup scenarios, where arbitrage is not executed within a single atomic transaction and there are delays between buy and sell operations introduce some risk. They found that price differences on rollups often last for 10–20 blocks, and discovered over 0.5 million unexploited arbitrage opportunities on these rollups.

In the rollup setting, Ferreira et al.~\cite{ferreira2024rolling} exploit the centralized sequencing mechanism of rollups to manipulate transaction ordering and mount cross-layer sandwich attacks. Their analysis estimates that attackers could have earned roughly 2 million USD through such sequencer-level manipulation. However, this line of work is fundamentally different from ours. Cross-layer attacks rely on privileged control or structural properties of a specific rollup’s sequencer, whereas we uncover an attack vector that arises purely from cross-chain message flow. Prior studies do not examine the possibility that cross-chain infrastructure itself can leak actionable transaction information and enable sandwich attacks across chains. To the best of our knowledge, our work is the first to identify, formalize, and empirically analyze cross-chain sandwich attacks, supported by real-world historical data.

\section{Conclusion and Future Works}
In this paper, we identify a previously unrecognized cross-chain sandwich attack in which an attacker exploits information leaked by cross-chain protocols to execute profitable front-running and back-running transactions. In this setting, existing defenses against sandwich attacks fail, allowing attackers to consistently outperform current MEV bots. We provide a theoretical analysis of the cross-chain sandwich attack and conduct an empirical study on historical data from a major cross-chain protocol. Our results show that this class of attacks produced at least \(0.56\) million USD in profit. To the best of our knowledge, we are the first to construct and publicly release a large-scale dataset of cross-chain sandwich attacks, providing a foundation for future research on cross-chain security. This work highlights a critical security risk introduced by cross-chain interoperability and aims to inspire more robust DeFi defense mechanisms.

\section{Acknowledgment}
This work is supported by the Guangzhou-HKUST(GZ) Joint Funding Program (No. 2024A03J0630 and No. 2025A03J3882), the Guangzhou Municipal Science and Technology Project (No. 2025A04J4168), and the Guangdong Provincial Key Lab of Integrated Communication, Sensing and Computation for Ubiquitous Internet of Things (No.2023B1212010007).

\bibliographystyle{plainurl}
\bibliography{ref}


\appendix

\begin{figure*}[t]
    \centering
    \includegraphics[width=0.95\linewidth]{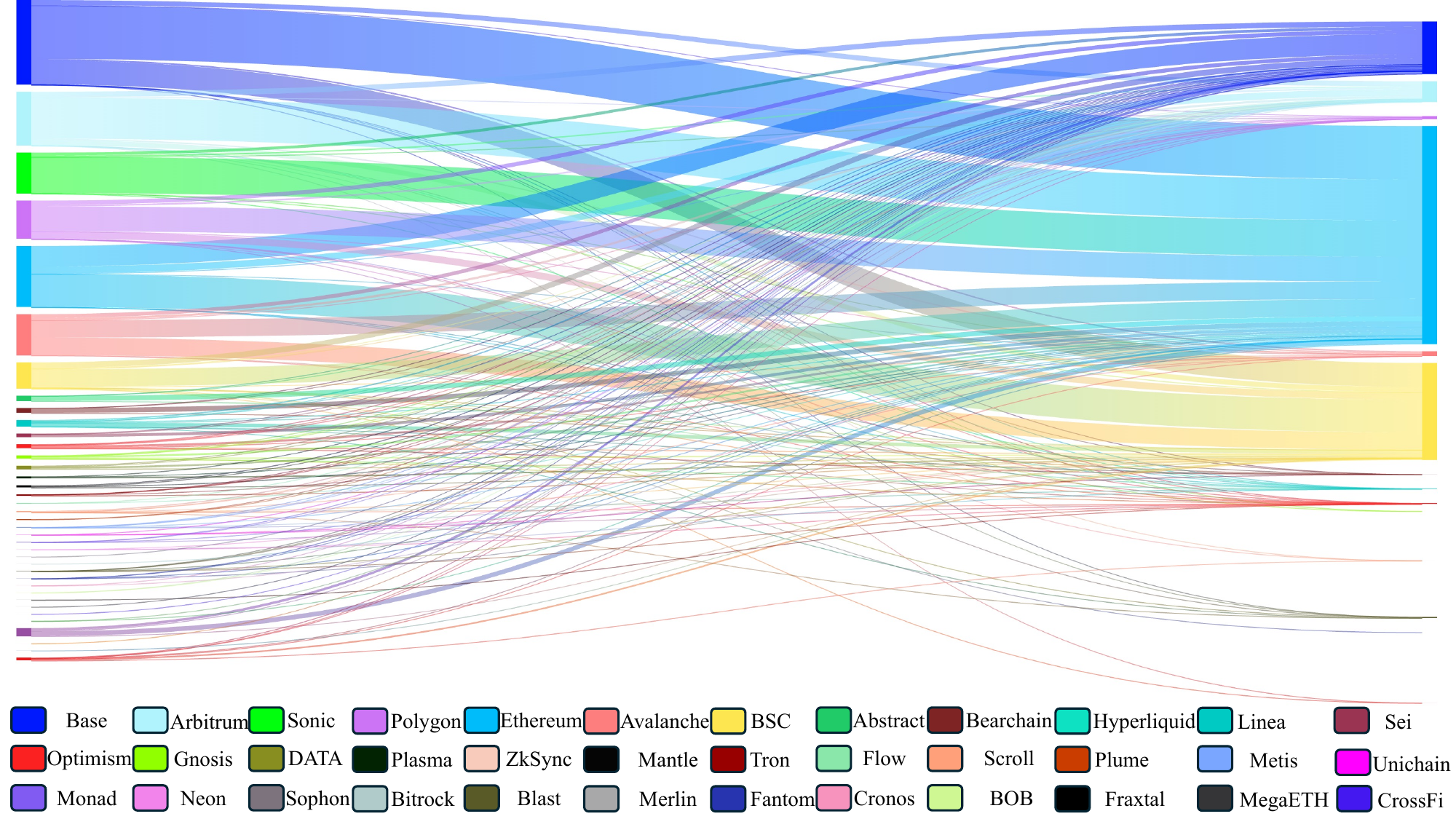}
    \caption{{Visualization of profits across different blockchains. The left nodes represent the source chains and the right nodes are the destination chains. The width of a node represents the attacker’s profit in USD realized on the destination chain.}}
    \label{sankey}
\end{figure*}

\section*{Related Data}
Table~\ref{exampleattack} lists some cross-chain sandwich attack pair examples. Table~\ref{events} presents an overview of the events and functions we leveraged in our detection.

\begin{table}[!htbp]
\caption{\textbf{Examples of cross-chain sandwich attacks}}
\label{exampleattack}
\centering
\scalebox{0.84}{
\begin{tabular}{llll}
\toprule
\(T_s\)&\(T_{A1}\)&\(T_v\)&\(T_{A2}\)\\
\midrule
\href{https://basescan.org/tx/0x3f48084945ac2ce89b787af9a0e719b4ecf33da7274c7974b886f236de00702d}{\textcolor{cyan}{0x3f48...702d}}&
\href{https://etherscan.io/tx/0x47048ec1fab14f923b218625b6597b25f41ed39e6a867185dfd3e514536a7f41}{\textcolor{cyan}{0x4704...7f41}}&
\href{https://etherscan.io/tx/0xaf7323c03ed04457e686601c79cc5f9e19c26cc4653bc138db25d27f6d5b6dc3}{\textcolor{cyan}{0xaf73...6dc3}}&
\href{https://etherscan.io/tx/0x3134c954eef067b72f377b5bca5b98622471d6f8a4f02b265a5cff00f6864e33}{\textcolor{cyan}{0x3134...4e33}}\\

\href{https://basescan.org/tx/0xe856e54cc70f41c0d0401f6dadd119c595513fee35bb1024d2da0667cadf384b}{\textcolor{cyan}{0xe856...384b}}&
\href{https://etherscan.io/tx/0x295f30e7376152fa1108991ace438b489308679f4dde555bc3026bcaef6c8b81}{\textcolor{cyan}{0x295f...8b81}}&
\href{https://etherscan.io/tx/0x913b3dde22e34bc4cc77ef756e89f8fc66635330efe31f88e9937749c56a3c47}{\textcolor{cyan}{0x913b...3c47}}&
\href{https://etherscan.io/tx/0x0ea7f589a86140433b2bbed470d18bd0ae0b1c468befe082c7f97590ff7ee17f}{\textcolor{cyan}{0x0ea7...e17f}}\\

\href{https://bscscan.com/tx/0x5f54eaca8b428abf47355b6e5bc6883e86eef94db71544aa7c52518d47dcfbfe}{\textcolor{cyan}{0x5f54...cfbe}}&
\href{https://etherscan.io/tx/0x7c7a451b79aebdad96be25d37270f849397cf32b644c0188a1528faabaf48cae}{\textcolor{cyan}{0x7c7a...48ca}}&
\href{https://etherscan.io/tx/0x0d56e906725eb41557b3452d45c56b9a583d13239d543044e2e034e7fbe3e0f7}{\textcolor{cyan}{0x0d56...e0f7}}&
\href{https://etherscan.io/tx/0xa886130242ae15dcc7121eaaa167f5a87c59142008325d3ed8508c4290e898be}{\textcolor{cyan}{0xa886...98be}}\\

\href{https://basescan.org/tx/0x904f8a8ed5d6137b66bd48e24c7ea7845fcf73524ea8578ab4c63b67edac1271}{\textcolor{cyan}{0x904f...1271}}&
\href{https://arbiscan.io/tx/0x56249f3b8bf21e7b8492637449e616eac80c3a78a0197729df4eade4a8a611d3}{\textcolor{cyan}{0x5624...611d3}}&
\href{https://arbiscan.io/tx/0x8d857f597ec2b076850b057ae445887928c2cf86acab86e7a88739872e3b77c5}{\textcolor{cyan}{0x8d85...77c5}}&
\href{https://arbiscan.io/tx/0x6552a1d55510c8f055585ff86560a58dd52c6cce8a778160bf21783ac8dec4dc}{\textcolor{cyan}{0x6552...dec4}}\\

\href{https://etherscan.io/tx/0x14ecbe817f34e6030287d604e8983f3215c0c602aee99850299aeed4d11282e7}{\textcolor{cyan}{0x14ec...82e7}}&
\href{https://basescan.org/tx/0x95e0224db8afc3102f50ef5ffb7ee32d601be2f1a631f85f7620e78a81dbaf92}{\textcolor{cyan}{0x95e0...af92}}&
\href{https://basescan.org/tx/0xe710365639b67a3c47f8ec01e86f513b45b90e2d22fb40f770632e951c8e57d2}{\textcolor{cyan}{0xe710...57d2}}&
\href{https://basescan.org/tx/0x0e2edfcc2d5ab1304efbe80782100ba5e04e8dec62ed528d2e485c995b4156aa}{\textcolor{cyan}{0x0e2e...56aa}}\\

\href{https://etherscan.io/tx/0xec1834e371a474b6a6e93fb3f99d3ba4c12443515e07cb7b708d5c63e09a0e3f}{\textcolor{cyan}{0xec18...0e3f}}&
\href{https://basescan.org/tx/0x244381c02a4171d01286254ac7966bd46136cb2f1ec0ac01b531c0e53327688d}{\textcolor{cyan}{0x2443...688d}}&
\href{https://basescan.org/tx/0x60cd0e20237931cc75695fbe15cde5a57450a07aa6e1c59b7e82ccae30af309c}{\textcolor{cyan}{0x60cd...309c}}&
\href{https://basescan.org/tx/0x56c172600acfbd57d434bef698d137fa4cc5dcd3d4b6e6faaad0fa6f4b872326}{\textcolor{cyan}{0x56c1...2326}}\\

\href{https://etherscan.io/tx/0x5b0b1f5063f30ebec7166334e607cb75257e01791ee0776d94339f5f90142c68}{\textcolor{cyan}{0x5b0b...2c68}}&
\href{https://basescan.org/tx/0x9b149a0b460bce3b805ac37162f1b5481c093c66c3efb2b6a185603094ac7a81}{\textcolor{cyan}{0x9b14...7a81}}&
\href{https://basescan.org/tx/0x5781e629ac4df9725ed9760220512f959c367652b975b4be5b254dd4376ccaa2}{\textcolor{cyan}{0x5781...caa2}}&
\href{https://basescan.org/tx/0xeed5bd78e87d3df91478f376753454ec44d2adda3110bc391703892d811e9973}{\textcolor{cyan}{0xeed5...9973}}\\

\href{https://etherscan.io/tx/0x57f6d34c180fd97e57ee22bd350acf836570ac0e0c2299799a76e283d73e45f2}{\textcolor{cyan}{0x57f6...e45f}}&
\href{https://basescan.org/tx/0x438aa944848bf388142bbb80fdf3bfd31a34383a66e73e51e17dbb0e9c6fd2e4}{\textcolor{cyan}{0x438a...d2e4}}&
\href{https://basescan.org/tx/0xb5651674397c4a99c31999281430f42995982b29604a44d0b0ffad818939dee7}{\textcolor{cyan}{0xb565...dee7}}&
\href{https://basescan.org/tx/0x88c48c87b25af062da1eb768a31c06c64d1c97dcd6f3953970cda3f007609585}{\textcolor{cyan}{0x88c4...9585}}\\

\href{https://etherscan.io/tx/0xdda50b3c1e8a399f9be86d5194da10fb9091f6a4390f276f9dd623942222bdd6}{\textcolor{cyan}{0xdda5...bdd6}}&
\href{https://bscscan.com/tx/0x07caa356b5f418dce46586ac55d1414328fad3c3b04bea0d9e00603ae95dc011}{\textcolor{cyan}{0x07ca...c011}}&
\href{https://bscscan.com/tx/0x5c5b604f49e3f2fc70bb20cce03da691bc74b2d06cc7561c29899191e8bfa0d5}{\textcolor{cyan}{0x5c5b...fa0d5}}&
\href{https://bscscan.com/tx/0x83d771c6e206c4a211857832db57f88a47da5a84a87ff16c8f09ea09aa1a25f5}{\textcolor{cyan}{0x83d7...25f5}}\\

\href{https://snowtrace.io/tx/0xac4a08843a93f963fb25f0a3f101c935d93465bdc7ba214ec4282cef88294153}{\textcolor{cyan}{0xac4a...4153}}&
\href{https://etherscan.io/tx/0x306e288ade3549a1eaf3684e483e0982ea26cff30c87bb0cb0372428c984ecf0}{\textcolor{cyan}{0x306e...ecf0}}&
\href{https://etherscan.io/tx/0x1aa21c8307f91ac9a2bc82a47a484d054b2a53799006384468f3bef9fca5182c}{\textcolor{cyan}{0x1aa2...182c}}&
\href{https://etherscan.io/tx/0xb134f5a04da7ee5abbd7b25452139a939f22dbf91f3a14649f2faa2bb9c6d205}{\textcolor{cyan}{0xb134...d205}}\\

\href{https://snowtrace.io/tx/0xf8e6acaeb707a49a72edf517cb3b582852b672e50eb35e594288f98de1564f00}{\textcolor{cyan}{0xf8e6...4f00}}&
\href{https://arbiscan.io/tx/0xf9bdef933dc133570da89a4833911983ae33551f54d4ff46543f224316363e08}{\textcolor{cyan}{0xf9bd...3e08}}&
\href{https://arbiscan.io/tx/0x56a841956c67f0c44c5dd8df31894b944e5530e85eeda6ecc762967f1fc5a34f}{\textcolor{cyan}{0x56a8...a34f}}&
\href{https://arbiscan.io/tx/0x2814daed6c4a37f7f73c3789fa4a71463b3777c4235427edbdede239933e466e}{\textcolor{cyan}{0x2814...e466e}}\\

\href{https://etherscan.io/tx/0xb8c580eda4e25802e43a9a86b46277a6c2c3f177a69bb14f8f41b2be5773f917}{\textcolor{cyan}{0xb8c5...f917}}&
\href{https://basescan.org/tx/0x82e8343c8bcb1a0ba15c34b813eeef72ebee3857073444de6fc8511dbce6dbe5}{\textcolor{cyan}{0x82e8...dbe5}}&
\href{https://basescan.org/tx/0xadcc9c7beb93097172c35f4fc6e9709c307452c052e47e748b12112a51cb7c63}{\textcolor{cyan}{0xadcc...7c63}}&
\href{https://basescan.org/tx/0xb1362f522ffc6035894c9858f1caddb8e96626607a561ed6a9815fa491175c69}{\textcolor{cyan}{0xb136...5c69}}\\

\href{https://ftmscan.com/tx/0x19a0ad4288b9cdfc212a39bd178e6b44d1ef2738641bec9eb81e2d80ba367abc}{\textcolor{cyan}{0x19a0...7abc}}&
\href{https://arbiscan.io/tx/0xaff74ac242a7ec8ef8f1c4818996ca455c5410d5d1554658d782f87d9399f526}{\textcolor{cyan}{0xaff7...f526}}&
\href{https://arbiscan.io/tx/0x5b7982a0d57f74d01bc5a2c9044311c9aa5fdf5a5b4d6db11967429cd95ac1d4}{\textcolor{cyan}{0x5b79...c1d4}}&
\href{https://arbiscan.io/tx/0xad85ee68512251e7bd0c931055414294542a15612649bf4767b582ad79750979}{\textcolor{cyan}{0xad85...0979}}\\

\href{https://etherscan.io/tx/0x2a9a288496693d4946f664b7f85f52c1810df162f4d8b05c4d6b4d4096767fcc}{\textcolor{cyan}{0x2a9a...7fcc}}&
\href{https://bscscan.com/tx/0xd4a2877923897debcd7b091c18a871e4ae04c269d96b71eec73cf3bb0abe1ea2}{\textcolor{cyan}{0xd4a2...1ea2}}&
\href{https://bscscan.com/tx/0x2a2ba8858f13f0ce8570272abbf0690534e89d254ce7bfd5bf3cda564403cd11}{\textcolor{cyan}{0x2a2b...cd11}}&
\href{https://bscscan.com/tx/0x03991c72fb50ef2feaff9de36066819ae6a96e47c10a782bf5ea79eca2657811}{\textcolor{cyan}{0x0399...7811}}\\

\href{https://bscscan.com/tx/0xa4695b809fa982acb31ac0ae8e8e575764f6d5a90b72028ccb4fe211c7bddf49}{\textcolor{cyan}{0xa469...df49}}&
\href{https://basescan.org/tx/0x9aefdbf7d74b01d6b76542ce5392cbd8eaca710bf013f5ed54bfe5f2ff4407ab}{\textcolor{cyan}{0x9aef...07ab}}&
\href{https://basescan.org/tx/0xb6aaeafc43ea60168635faf43141a541498436ead69fa635bbaa805dbc94d09e}{\textcolor{cyan}{0xb6aa...d09e}}&
\href{https://basescan.org/tx/0xae378143e9a93b16715c00fdc591a8c8105fdd65d29500018054174009d3ea87}{\textcolor{cyan}{0xae37...ea87}}\\

\href{https://basescan.org/tx/0x20967fee57053fe1cff093e502a01a352383e35b7aa48ebdcb397b0f8653b067}{\textcolor{cyan}{0x2096...b067}}&
\href{https://arbiscan.io/tx/0x5791637c3d2e48470624dc2a43c269441797e73b4147be479e794243d1a6547c}{\textcolor{cyan}{0x5791...547c}}&
\href{https://arbiscan.io/tx/0x92314b6a15736594318404ce6508e772cb4eac5a9f641685839e186ea57c1547}{\textcolor{cyan}{0x9231...c547}}&
\href{https://arbiscan.io/tx/0x6bb58e316273aeda7ead47d3e537de6e9296e9d6512d57cfaf4523e8fc4b2478}{\textcolor{cyan}{0x6bb5...2478}}\\

\href{https://etherscan.io/tx/0x2a7f0f79cdfc865b276d50d7d56234bd90ceea67ce8a9e87ab5af9d6b77104d0}{\textcolor{cyan}{0x2a7f...104d}}&
\href{https://arbiscan.io/tx/0x6b65a32437ccfec139b5a229bb9fbb7d932de7dd31a461e74972c857bebaf57a}{\textcolor{cyan}{0x6b65...af57}}&
\href{https://arbiscan.io/tx/0xe0876e3518766205b851a9b22f9208fe00f9761a0374d5cf5b4303a8a301cf6b}{\textcolor{cyan}{0xe087...1cf6}}&
\href{https://arbiscan.io/tx/0xdbe8977dcba4cd2c8283866ebda6c815ee623a96981da628856d4df72cd1a6ae}{\textcolor{cyan}{0xdbe8...a6ae}}\\

\href{https://basescan.org/tx/0x8883d4c755a509b26df64904704dda3c97494283a299051610151fc49e84ee32}{\textcolor{cyan}{0x8883...ee32}}&
\href{https://bscscan.com/tx/0xfe43bcfb2f35433f4397c4d3a2310af3aaa2257cf349c01edc18075761050624}{\textcolor{cyan}{0xfe43...0624}}&
\href{https://bscscan.com/tx/0xadbbb5cd6af608911a978cec82ff8bdafbc5aa99ec22443300fb51c2296c9050}{\textcolor{cyan}{0xadbb...9050}}&
\href{https://bscscan.com/tx/0xf1093836a25e355acbf82bfaf1cda0535ef371d105e654e176533bc515e78bfb}{\textcolor{cyan}{0xf109...8bfb}}\\

\href{https://etherscan.io/tx/0x55aa29c8b7f3d15416fff6a8079c4efd7f698cb5efbb5bc9767d68b3a584d10b}{\textcolor{cyan}{0x55aa...d10b}}&
\href{https://lineascan.build/tx/0xe7e3f54ea8c374d88c51199b423941821e5881036a1c88a41fd05b621c3d8410}{\textcolor{cyan}{0xe7e3...8410}}&
\href{https://lineascan.build/tx/0xd31ee445ad2b5eb4555e6c21727a4c8526dff2b8441f5f535269a4b61608aecc}{\textcolor{cyan}{0xd31e...aecc}}&
\href{https://lineascan.build/tx/0xc465ceaa3a445d1517378872a8881cee98e1e79e8c0a3909d351cfeed70265bf}{\textcolor{cyan}{0xc465...65bf}}\\

\href{https://arbiscan.io/tx/0x7f0539f48eb3a7da39f4316097207df4b510c71d8fa060d955b5a9c9d849425b}{\textcolor{cyan}{0x7f05...425b}}&
\href{https://basescan.org/tx/0xb3c9171df045c282b2d39ca4d8703e8af1690ba9a75a6bfa503f75e750c33f66}{\textcolor{cyan}{0xb3c9...3f66}}&
\href{https://basescan.org/tx/0xbd1ae537b318da1395b0a0c0b61b395898311ee1f0da50f7c9dfa05a2bf2afe6}{\textcolor{cyan}{0xbd1a...afe6}}&
\href{https://basescan.org/tx/0x0374b108cd06324c9fa134e3928b375017077207ebf540d32497cf2b794adf94}{\textcolor{cyan}{0x0374...adf94}}\\

\bottomrule
\end{tabular}
}
\end{table}

\begin{table*}[!htbp]
\caption{\textbf{Pools sorted by numbers of attacks}}
\label{pooltims}
\centering
\scalebox{0.9}{
\begin{tabular}{lcclr}
\toprule
Address&Tokens&Chain&Protocol& Number of attacks\\
\midrule
0x172fcd41e0913e95784454622d1c3724f546f849&BUSD-WBNB&BSC&PancakeSwap V3&2473\\
0x72ab388e2e2f6facef59e3c3fa2c4e29011c2d38&USDC-WETH&Base&PancakeSwap V3&2351 \\
0xf2688fb5b81049dfb7703ada5e770543770612c4&USDC-WBNB&BSC&PancakeSwap V3&2170  \\
0x47a90a2d92a8367a91efa1906bfc8c1e05bf10c4&BUSD-WBNB&BSC&Uniswap V3&1489 \\
0x6f38e884725a116c9c7fbf208e79fe8828a2595f&USDC-WETH&Arbitrum&Uniswap V3&863 \\
0xe0554a476a092703abdb3ef35c80e0d76d32939f&USDC-WETH&Ethereum&Uniswap V3&805 \\
0xb4cb800910b228ed3d0834cf79d697127bbb00e5&USDC-WETH&Base&Uniswap V3&661\\
0x62fcb3c1794fb95bd8b1a97f6ad5d8a7e4943a1e&WBNB-ETH&BSC&PancakeSwap V3&501 \\
0xdbc6998296caa1652a810dc8d3baf4a8294330f1&USDC-WETH&Base&Aerodrome&442\\
0x7fcdc35463e3770c2fb992716cd070b63540b947&USDC-WETH&Arbitrum&PancakeSwap V3&419\\
0xf6c0a374a483101e04ef5f7ac9bd15d9142bac95&USDBC-WETH&Base&PancakeSwap V3&403\\
0xb1026b8e7276e7ac75410f1fcbbe21796e8f7526&USDC-WETH&Arbitrum&Algebra&324\\
0x9f599f3d64a9d99ea21e68127bb6ce99f893da61&BUSD-ETH&BSC&PancakeSwap V3&312  \\
0x482fe995c4a52bc79271ab29a53591363ee30a89&USDC-WETH&Base&Uniswap V3&290\\
0xc7bbec68d12a0d1830360f8ec58fa599ba1b0e9b&USDT-WETH&Ethereum&Uniswap V3&236\\
0x4141325bac36affe9db165e854982230a14e6d48&USDC-WBNB&BSC&Uniswap V3&221\\
0x86f1d8390222a3691c28938ec7404a1661e618e0&WPOL-WETH&Polygon&Uniswap V3&217\\
0x389938cf14be379217570d8e4619e51fbdafaa21&USDT0-WETH&Arbitrum&PancakeSwap V3&177\\
0xb6e57ed85c4c9dbfef2a68711e9d6f36c56e0fcb&USDC-WPOL&Polygon&Uniswap V3&134\\
0x5968feacba91d55010975e0cfe8acfc32664ad33&BUSD-ZKJ&BSC&PancakeSwap V3&127\\
0xb94b22332abf5f89877a14cc88f2abc48c34b3df&USDC-cbBTC&Base&PancakeSwap V3&121\\
0xc6962004f452be9203591991d15f6b388e09e8d0&USDC-WETH&Arbitrum&Uniswap V3&116\\
0xfae3f424a0a47706811521e3ee268f00cfb5c45e&USDC-WAVAX&Avalanche&Uniswap V3&115\\
0x88e6a0c2ddd26feeb64f039a2c41296fcb3f5640&USDC-WETH&Ethereum&Uniswap V3&105\\
0xa20c959b19f114e9c2d81547734cdc1110bd773d&USDC-WAVAX&Avalanche&Aerodrome&103\\
0xe30d5bf485f7476ac15884a28ffb3c9cea635dcb&USDC-AVNT&Base&Aerodrome&103\\
0x62cf00528cb7af872c1f9dd426e655c903f16770&USDC-ETH&BSC&PancakeSwap V3&101  \\
0xc211e1f853a898bd1302385ccde55f33a8c4b3f3&WETH-cbBTC&Base&PancakeSwap V3&100\\
\bottomrule
\end{tabular}
}
\end{table*}

\begin{table*}
\centering
\refstepcounter{table}
\label{events}
{\footnotesize TABLE~\thetable\\
\textbf{OVERVIEW OF EVENTS AND FUNCTION WE USED IN THE DETECTION}\par}
\vspace{0.5em}
\resizebox{\textwidth}{!}{
\begin{tabular}{lll}
\toprule
\textbf{Protocol}&\textbf{Event Name}&\textbf{Event Topic Hash}\\
\midrule
Uniswap V2&Swap&0xd78ad95fa46c994b6551d0da85fc275fe613ce37657fb8d5e3d130840159d822\\
PancakeSwap V2&Swap&0xd78ad95fa46c994b6551d0da85fc275fe613ce37657fb8d5e3d130840159d822\\
Uniswap V3&Swap&0xc42079f94a6350d7e6235f29174924f928cc2ac818eb64fed8004e115fbcca67\\
PancakeSwap V3&Swap&0x19b47279256b2a23a1665c810c8d55a1758940ee09377d4f8d26497a3577dc83\\
Symbiosis&OracleRequest&0x532dbb6d061eee97ab4370060f60ede10b3dc361cc1214c07ae5e34dd86e6aaf\\
DeBridge&CreatedOrder&0xfc8703fd57380f9dd234a89dce51333782d49c5902f307b02f03e014d18fe471\\
\bottomrule
\textbf{Protocol}&\textbf{Function Name}&\textbf{Function Signature}\\
\midrule
Symbiosis&metaMintSyntheticToken&0xc29a91bc\\
Symbiosis&metaBurnSyntheticToken&0xe66bb550\\
Symbiosis&receiveRequestV2Signed&0x84d61c97\\
Symbiosis&metaUnsynthesize&0xc23a4c88\\
Symbiosis&externalCall&0xf5b697a5\\
DeBridge&strictlySwapAndCallDln&0xc7a76969\\
1inch&swap&0x12aa3caf\\
1inch&uniswapV3SwapTo&0xbc80f1a8\\
OpenOcean&swap&0x90411a32\\
Uniswap V2&swapExactTokensForTokens&0x38ed1739\\
Uniswap V2&swapExactTokensForETH&0x18cbafe5\\
\bottomrule
\end{tabular}
}
\end{table*}

\section*{Ethics Considerations}
We conduct this study with attention to ethical risks and responsible disclosure. We do not perform any active attacks on Mainnet or test nets to validate or reproduce cross-chain sandwich attacks and we explicitly discourage replication of such attack on the blockchain Mainnet. All of our analysis is based on historical, publicly available on-chain data. To facilitate future research while managing potential misuse risks, we have only released the dataset we collected, which is intended solely for academic use. At the same time, to prevent the attack from being replicated or misused, we do not release the corresponding code.

We evaluate the potential harms of such cross-chain sandwich attack and included concrete mitigation suggestions in the paper to help reduce the real-world risk posed by the vulnerabilities we describe. 
Prior to public release, we notified the development team of Symbiosis, deBridge and other related protocols, and provided them with a technical summary of our findings and recommended mitigations.

We commit to follow-up work to assess the effectiveness of proposed mitigations and to update our dataset and recommendations as protocols evolve. Where we provide additional or more detailed artifacts in the future, we will do so under the same controlled-access process and principles.

\section*{Open Science}
Our dataset is open sourced on Kaggle\footnote{https://www.kaggle.com/datasets/anonymous8397458/cross-chain-sandwich-datasets}. To enable independent verification and further research, we make available 
the processed cross-chain transaction datasets and the corresponding records 
generated by our sandwich attack detection algorithm. For Symbiosis, we 
collected historical cross-chain transaction data through the API exposed by 
its public explorer\footnote{https://api.symbiosis.finance/explorer/v2/transactions}, covering finalized cross-chain swaps between November 21, 
2022 and April 4, 2026. For deBridge, we collaborated with the protocol team 
and obtained access to its historical transaction API\footnote{https://dln-api.debridge.finance/api/Orders/filteredList}, which allowed us to 
retrieve completed cross-chain orders from December 21, 2022 to April 4, 2026.

Each released source dataset describes a completed cross-chain transaction as 
a transaction pair $(T_s,T_v)$, where $T_s$ denotes the initiating transaction 
on the source chain and $T_v$ represents the corresponding execution 
transaction on the destination chain. Symbiosis records contain the essential 
transaction-pair information and execution timestamps, while deBridge records 
provide additional metadata, including order IDs, source and destination chain 
information, token addresses and symbols, token decimals, and transfer amounts 
in both raw and processed formats.

In addition to the original cross-chain transaction records, we provide the 
outputs of our sandwich attack analysis pipeline. These records include the 
identified victim transactions, targeted liquidity pools, candidate attacker 
swaps occurring before and after the victim transaction, and the final matched 
sandwich patterns represented as $(T_{A1},T_v,T_{A2})$. This additional 
information enables researchers to reproduce our detection process and 
validate the reported measurement results.

\end{document}